\documentclass[onecolumn,amsmath,amssymb,12pt,superscriptaddress,nofootinbib,floatfix]{revtex4}
\pdfoutput=1

\usepackage[latin1]{inputenc}
\usepackage[english]{babel}
\usepackage{amssymb}
\usepackage{amsmath}
\usepackage{amsthm}
\usepackage[]{graphicx}
\usepackage[]{subfigure}
\usepackage{tensor}
\usepackage{color}
\usepackage{cancel}
\usepackage{setspace}
\usepackage{fancyhdr}
\usepackage[bookmarks,linktocpage, colorlinks=true, plainpages = false, citecolor = blue,  linkcolor=blue, urlcolor = blue, filecolor = blue]{hyperref}

\def\1p{{(1p)}}

\def\p0{\phi_0}

\def\be{\begin{equation}}
\def\ee{\end{equation}}
\def\beq{\begin{eqnarray}}
\def\eeq{\end{eqnarray}}

\newcommand{\pkt}{\; .}
\newcommand{\kma}{\; ,}

\def\e{{\rm e}}

\begin{document}

\title{Scalar Lumps with Two Horizons}

\author{George Lavrelashvili}
\email[]{george.lavrelashvili@tsu.ge}
\affiliation{Department of Theoretical Physics, A.Razmadze Mathematical Institute \\
	at I.Javakhishvili Tbilisi State University, GE-0186 Tbilisi, Georgia}
\author{Jean-Luc Lehners}
\email[]{jlehners@aei.mpg.de}
\affiliation{Max Planck Institute for Gravitational Physics \\ (Albert Einstein Institute), 14476 Potsdam-Golm, Germany}


\begin{abstract}
\vspace{1cm}
\noindent
We study generalisations of the Schwarzschild-de Sitter solution in the presence of a scalar field with a potential barrier. These static, spherically symmetric solutions have two horizons, in between which the scalar interpolates at least once across the potential barrier, thus developing a lump. In part, we recover solutions discussed earlier in the literature and for those we clarify their properties. But we also find a new class of solutions in which the scalar lump curves the spacetime sufficiently strongly so as to change the nature of the erstwhile cosmological horizon into an additional trapped horizon, resulting in a scalar lump surrounded by two black holes. These new solutions appear in a wide range of the parameter space of the potential. We also discuss (challenges for) the application of all of these solutions to black hole seeded vacuum decay.
\end{abstract}
\maketitle
\newpage
\tableofcontents

\section{Introduction}

The no-hair conjecture, formulated as no-hair theorems for concrete matter contents \cite{Chrusciel:2012jk}, restricts the ways one can attach nontrivial field configuration to a black hole solution. This topic received renewed actuality since the detection of gravitational waves from black hole collisions not only demonstrated the existence of black holes, but will also allow for increasingly accurate measurements of their properties. One of the ways to overcome the restrictions of the no-hair conjecture is to consider a self-interacting scalar field theory with non-trivial potential, and coupled to gravity. It was shown in \cite{Torii:1998ir} that in the case of a non-monotonic scalar field potential there are static Schwarzschild-de Sitter (SdS) like solutions with the scalar field oscillating between different sides of a potential barrier.
Such solutions with two horizons and an oscillating scalar field were recently rediscovered in \cite{Gregory:2020cvy} in a completely different context, namely that of Hawking-Moss cosmological phase transitions. In \cite{Gregory:2020cvy} only the properties of solutions in between the two horizons were investigated, since this is the only part that is relevant for phase transitions described by Euclidean methods.

The main aim of our study is to perform a general investigation of the properties of such static, spherically symmetric oscillating solutions with two horizons in a scalar field theory minimally coupled to gravity.\footnote{Solutions with a regular origin surrounded by a single horizon are closely related, and were investigated in \cite{Torii:1999uv,Lavrelashvili:2021rxw}.} We call such solutions ``scalar lumps''. For definiteness, and ease of comparison with earlier works, we choose the potential to be a double well, supplemented by a positive cosmological constant. Our results partly overlap with the findings of \cite{Torii:1998ir} and \cite{Gregory:2020cvy}, but we disagree in several important points. Our three main findings are:

First, we have discovered two entirely new classes of solutions. The metric ansatz used in earlier works had a spherical part just as in the Schwarzschild metric, $ds^2 \supset r^2 d\Omega^2,$ for which the sphere radius grows monotonically with the coordinate $r.$ As already suggested by the Nariai limit of the Schwarzschild-de Sitter solution, in which the sphere radius is constant, this is too restrictive. Allowing the radius to be a function $R(r)$, i.e. $ds^2 \supset R(r)^2 d\Omega^2,$ leads to two new types of solutions. In one class $R(r)$ starts shrinking again beyond the cosmological horizon, which indicates that the horizon shields a re-collapsing universe. In a second new class, the sphere radius $R(r)$ already turns around in between the two horizons, due to the strong gravitational effect of the scalar lump. In the latter case, the nature of the second horizon changes and it turns into a second black hole horizon. We have thus discovered scalar lump solutions that are surrounded by two black holes (in general of different sizes) in opposite radial directions! Let us also mention that, for a fixed potential, when solutions exist they arise as a one-parameter family, where the parameter in question represents the size of one of the horizons. This is in direct analogy with the SdS solution. However, here one has the additional feature that for different horizon sizes the nature of the solutions may switch between the three types described above.

Second, we find that the three types of solutions exist for wide ranges of parameters of the potential (now given a fixed size of one horizon), but the precise limits where solutions cease to exist differ from those presented in earlier work \cite{Torii:1998ir}. There are several reasons for this discrepancy: one is certainly the increased ease and speed of numerical calculations that is nowadays possible. Another is that we have come to realise that there are different types of limiting solutions, not just the SdS solution sitting at the top of the potential barrier, but also the Nariai solution, and combinations of the two that only exist because of our better adapted metric ansatz. This leads to more intricate domains of existence of the various types of solutions.

Third, we have re-examined some aspects of vacuum decay seeded by black holes. In between the two horizons, the solutions we are describing are time independent, and they may thus be rather trivially analytically continued to Euclidean time (modulo the choice of periodicity in the Euclidean time direction). In order to describe vacuum decay one is instructed to glue such a scalar lump solution at one of its horizons to the cosmological horizon of a surrounding SdS spacetime. An analysis of the boundary conditions of scalar lump solutions shows that such a gluing may however not be done in a smooth manner (contrary to the assumptions in \cite{Gregory:2020cvy}). We will discuss possible interpretations/resolutions of this obstruction.


\section{Ansatz and field equations}

We will consider gravity minimally coupled to a scalar field $\varphi$ with a potential $V(\varphi)$ that we will restrict to be positive. The action is given by
\begin{align}
S=\int d^4 x \sqrt{-g} \big[\frac{1}{2\kappa}{\mathcal R}- \frac{1}{2}g^{\mu\nu}\partial_\mu\varphi\partial_\nu\varphi - V(\varphi) \big] \kma
\end{align}
where $\kappa = 8 \pi G$.
In what follows we will assume spherical symmetry. A general spherically symmetric metric can be written as \cite{Landau}
\begin{align}
ds^2= - l(r, t) dt^2 + a(r, t)drdt +h(r, t) dr^2 + k(r, t) d\Omega^2_2 \kma
\end{align}
where $d\Omega^2_2=d\theta+sin^2 \theta d\phi^2$.
Within this ansatz we are still free to perform coordinate transformations which respect spherical symmetry,
i.e. we can change from the variables $(r, t)$ to new variables $(\tilde{r}, \tilde{t})$, such that
\begin{align}
r=f_1 (\tilde{r}, \tilde{t}),~~~~t=f_2 (\tilde{r}, \tilde{t}) \kma
\end{align}
where $f_1$ and $f_2$ are arbitrary functions.
Using this freedom one can eliminate the non-diagonal terms, $a(r, t)=0$ and e.g.
set  $k(r, t) = r^2$, (Schwarzschild gauge). This is possible if $\frac{dk(r, t)}{dr} \neq 0$.
Another convenient choice is $l(r, t)= h(r, t)^{-1}$.

In what follows we will be interested in static, spherically symmetric fields and
parameterise the metric  as
\begin{align}\label{eq:generalgauge}
ds^2= - f(r) \e^{2\delta(r)} dt^2 + \frac{dr^2}{f(r)} +R^2(r) d\Omega^2_2 \kma
\end{align}
In this parametrisation the reduced action takes the form
\begin{align} \label{eq:redact}
S^{\rm red} = 4 \pi T \int dr \e^{\delta} \big[ \frac{1}{\kappa} \big( 1+f \{ {R'}^2+(\delta' +\frac{f'}{2f}) (R^2)' \} \big)
-\frac{1}{2}f R^2 {\varphi'}^2 - R^2 V \big] \kma
\end{align}
where $' \equiv d/dr$ and $T=\int dt$ is an overall constant stemming from the integration over the time coordinate. 
We can now discuss two convenient gauges.

\subsection{Schwarzschild gauge}

As long as $R'(r)\neq 0$, we can choose Schwarzschild gauge $R(r) \equiv r$ .
Varying the reduced action Eq.(\ref{eq:redact}) w.r.t. $\varphi, \delta$ and $f$
one obtains the equations of motions in this gauge:
\begin{align}
\varphi'' &= -(\frac{2}{r}+\frac{f'}{f}+\delta') \varphi'+ \frac{1}{f}\frac{\partial V}{\partial \varphi} \kma \label{eq:phiequation} \\
rf' &=1-f- \kappa r^2 (\frac{1}{2}f \varphi'^2 + V)\kma \label{eq:fequation}\\
\delta' &=\frac{\kappa}{2}  r \varphi'^2 \pkt \label{eq:dequation}
\end{align}
Introducing a ``mass'' function $\mu (r)$ via the  relation
$ f \equiv 1 - \frac{2 \kappa \mu}{r},$
we get the following equation for $\mu,$
\begin{align} \label{eq:mupSdS}
\mu' =\frac{1}{2}  r^2 (\frac{1}{2}f \varphi'^2 + V) \pkt
\end{align}

As is clear from Eq.~(\ref{eq:dequation}), the metric function $\delta$ decouples from the main system.
Once we determine $\varphi (r)$ and $f(r)$ from the coupled system of Eqs.~(\ref{eq:phiequation}, \ref{eq:fequation}),
we can easily find the metric function $\delta$ by integrating  Eq.~(\ref{eq:dequation}).

Variation w.r.t. $R$ gives
\begin{align}
\frac{1}{2}f'' + f\delta''+ \frac{3}{2} \delta' f' +\frac{1}{r} f' + f \delta'^2+\frac{1}{r}f \delta' = - \kappa (\frac{1}{2}f \varphi'^2 +  V) \pkt
\end{align}
One can check that this equation is consequence of above equations of motion and does not give any new information.

We will be interested in generalised Schwarzschild-de Sitter (SdS) solutions where the metric function $f$ has two simple zeros,
two horizons. In the standard SdS case these are a black hole horizon and a cosmological horizon.
The equations of motion have regular singular points where $f$ has zeroes.
It is known that generally at such points not all solutions stay regular,
and one should analyse the behaviour of solutions close to these singular points in order to find a regular branch of solutions.
Close to a horizon at $r=r_h$ we obtain a two parameter family of regular solutions, which behave as
\begin{align}
\varphi(\rho) &= \varphi_h + \frac{r_h V'(\varphi_h)}{1-\kappa r_h^2 V(\varphi_h)} \rho + O(\rho^2) \kma \label{eq:phiTaylorBH}\\
f(\rho) &= \frac{1- \kappa r_h^2 V(\varphi_h)}{r_h} \rho  + O(\rho^2) \kma \label{eq:fTaylorBH}
\end{align}
where $\rho=r-r_{h}$ and $V' \equiv dV/d\varphi$. Here $\varphi_h$ and $r_h$ are the free parameters.

For the metric function $\delta$ we find
\begin{align}
\delta(\rho) &= \delta_h + \frac{\kappa r_h^3 V'(\varphi_h)^2}{2(1-\kappa r_h^2 V(\varphi_h))^2} \rho + O(\rho^2) \kma
\end{align}
where $\delta_h$ is a further free parameter which determines the normalisation of the $t$ coordinate.

\subsection{Schwarzschild-de Sitter solutions} \label{subSdS}

At an extremum $\varphi_m$ of the scalar field potential, $V'(\varphi_m)=0,$ the scalar may be constant and act as vacuum energy. The metric then takes the form of the Schwarzschild-de Sitter solution
\begin{align} \label{eq:SdS-feq}
f_{0}(r) &= 1 - \frac{2 \kappa M_0}{r} -\frac{\kappa V(\varphi_m)}{3} r^2 \\
\mu_{0}(r)&= M_0 + \frac{V(\varphi_m)}{6} r^3 \pkt \label{eq:muSdS}
\end{align}
This is a one-parameter family of solutions parameterised by $M_0,$ which is the analogue of the black hole mass in asymptotically flat space. Given $M_0$, one finds the two horizons $r_{0,b}$ and $r_{0,c}$ as the positive roots of the cubic equation
\begin{align}\label{eq:horizons}
f_{0}(r) &= 0 \pkt
\end{align}
The smaller root $r_{0,b}$ is interpreted as a black hole horizon. It corresponds to a trapped horizon and beyond it the metric is that of a crunching Kantowski-Sachs universe. The larger root $r_{0,c}$ is interpreted as a cosmological horizon. It is an anti-trapped surface beyond which we find a universe in accelerated expansion that asymptotes to de Sitter spacetime. In between the two horizons the solution is static.

As the ``mass'' $M_0$ is increased from zero, the black hole horizon $r_{0,b}$ also grows from zero size, while the cosmological horizon $r_{0,c}$ decreases from the maximum radius $\sqrt{\frac{3}{\kappa V(\varphi_m)}}.$ The two horizons become equal, $r_{0,b}=r_{0,c}=\frac{1}{\sqrt{\kappa V(\varphi_m)}},$ at the limiting mass $M_{N}=\frac{1}{\sqrt{9\kappa V(\varphi_m)}}.$ In this so-called Nariai limit, the radius of the two-sphere is in fact constant everywhere and the metric is given by
\begin{align}
ds^2= - f_N(r)dt^2 + \frac{dr^2}{f_N(r)} +\frac{1}{\kappa V(\varphi_m)} d\Omega^2_2 \kma \qquad f_N(r)=1-V(\varphi_m)r^2 \,.
\end{align}
This spacetime is the direct product of 2-dimensional de Sitter space and the 2-sphere. A curious feature of the Nariai limit is that the sphere radius abruptly becomes constant, and the Schwarzschild gauge inapplicable. This already suggests that in looking for generalisations, Schwarzschild gauge may not be the best choice. We further know from past experience with solutions with a single horizon that the sphere radius may well be non-monotonic. Hence it will be useful to use a gauge that allows for a general evolution of the sphere radius.

\subsection{General gauge}

If at some point $R'(r) = 0$ the Schwarzschild gauge becomes inappropriate. In such a case we may set $\delta \equiv 0$ while leaving $R(r)$ general, and we call this choice ``general'' gauge.
Varying the reduced action Eq.(\ref{eq:redact}) w.r.t. $\varphi, R$ and $f$ we obtain the corresponding field equations:
\begin{align}
\varphi'' &= -(2\frac{R'}{R}+\frac{f'}{f}) \varphi'+ \frac{1}{f}\frac{\partial V}{\partial \varphi} \kma \label{eq:phi}\\
f'' &= -2 \frac{R'}{R} f' - 2 \kappa V \kma \label{eq:f} \\
R'' &= - \frac{\kappa}{2} R \varphi'^2 \kma \label{eq:R}
\end{align}
while variation w.r.t. $\delta,$ prior to setting it to zero, gives the constraint equation
\begin{align}
f \frac{R'^2}{R^2 } + f' \frac{R'}{R} = \frac{1}{R^2 }+ \kappa (\frac{1}{2}f \varphi'^2 -V) \pkt \label{eq:constraint}
\end{align}
Hence we see that solutions may very well have $R'=0$. At such a point the constraint provides us with the value of the extremal radius
\begin{align}
\frac{1}{R^2} = \kappa \left( V(\varphi) - \frac{1}{2} f \varphi'^2\right)   \pkt
\end{align}

Close to a horizon where $f(r_h)=0,$ we can solve the equations of motion again as a Taylor series, revealing the behaviour of regular solutions on the horizon. This time we obtain a 4-parameter family of solutions specified by the expansions
\begin{align}
\varphi (\rho) &= \varphi_h + \frac{V'(\varphi_h)}{f'_{h}} \rho
+ \frac{V'(\varphi_h) [2 \kappa V(\varphi_h) +V''(\varphi_h)]}{4 {f'_{h}}^2}\rho^2 + O(\rho^3) \kma \label{eq:phiTaylorBHgg} \\
f(\rho) &= f'_{h} \rho - \frac{1}{R_h^2}\rho^2 + O(\rho^3) \kma \label{eq:fTaylorBHgg}\\
R(\rho) &= R_{h} + \frac{1-\kappa R_{h}^2 V(\varphi_h) }{f'_{h} R_{h}} \rho
- \frac{\kappa R_{h} V'(\varphi_h)^2}{4 {f'_{h}}^2}   \rho^2 + O(\rho^3) \kma \label{eq:RTaylorBHgg}
\end{align}
where $\rho\equiv r-r_{h}$ and $r_{h}, f'_{h},\varphi_{h}$ and $R_{h}$ are free parameters.

We will be interested in solutions with 2 horizons. The expansions will then be of analogous form at both horizons, but generally with different values of the parameters. When the scalar potential is positive, two is also the maximum number of horizons that are allowed. This can be seen from Eq. \eqref{eq:f} which implies that any extremum in $f$ is necessarily a maximum. Thus, for positive potentials, once $f$ has turned negative it cannot turn around again but will decrease towards minus infinity. Meanwhile the equation for $R$, Eq. \eqref{eq:R}, implies that $R$ always has a negative second derivative, given that $R$ is always positive. This leaves three cases: $R$ may be constant, as in the Nariai solution. 
Or $R$ may be a monotonic function and have a single zero, if the scalar field profile is not too prominent. Finally, the evolution of the scalar field may cause $R$ to develop a maximum, in which case $R(r)$ will develop two zeroes which are both generally speaking associated with a spacetime singularity.\footnote{With fine tuning of the parameters one can hope to find globally regular compact solutions, such as those found in a similar settings in the Einstein-Yang-Mills theory with positive cosmological constant \cite{Volkov:1996qj}. It was shown in \cite{Lavrelashvili:2021rxw} that in the scalar field theory under consideration this is only  possible if the potential minima are located at negative values of potential.}
This last possibility will feature prominently in our numerical results presented below. In all cases we find that cosmic censorship is respected, and that singularities are shielded by horizons.


\section{Numerical results}

The scalar potential that we will consider is of double well form,
\begin{align}
V(\varphi) =  \frac{\Lambda}{\kappa} + \frac{\lambda}{4}(\varphi^2-v^2)^2\,,
\end{align}
where we have also included a cosmological constant $\Lambda.$ This potential contains a local maximum at $\varphi=0,$ of magnitude $V(0)=\frac{\Lambda}{\kappa} + \frac{\lambda}{4}v^4,$ and two minima at $\varphi=\pm v.$
In what follows we will use $G=1$ units (implying $\kappa = 8\pi$).
Furthermore, as shown in \cite{Torii:1998ir}, the fields may be redefined so as to effectively set $\Lambda=1$. This implies that we do not have to consider different values for the cosmological constant in searching for solutions,
specifying only the parameters $\lambda, v$ of the potential is sufficient.

Our numerical strategy is as follows. In general gauge we have 3 functions, $f(r), R(r)$ and $\phi(r)$ satisfying second order equations, which implies that one needs to specify 6 parameters to fix a solution. The constraint equation Eq.~(\ref{eq:constraint}) reduces the number of free parameters by one, leaving one with only 5 free parameters required for specifying a solution. As shown above in Eqs.~(\ref{eq:phiTaylorBHgg}, \ref{eq:fTaylorBHgg}, \ref{eq:RTaylorBHgg}), close to a horizon we have 4 free parameters. At the first and second horizons we will respectively call these parameters $r_{1,2}, f'_{1,2}, \varphi_{1,2}$ and $R_{1,2},$ i.e. we have 8 parameters in total.
With our metric ansatz there actually remains some residual gauge freedom: note that equations of motions are translationally invariant w.r.t. a shift of the variable $r$,
\begin{align}
r \to r + \Delta \kma
\end{align}
and scale invariant w.r.t. the following transformations
\begin{align}
r \to \alpha r,~~~~f \to \alpha^2 f \kma
\end{align}
with arbitrary parameters $\Delta$ and $\alpha$. These transformations allow us to fix the coordinate location of the first horizon to be at the origin, $r_1=0,$ and they also allow us to fix the derivative of $f$ at the location of the first horizon -- we will set $f_1^\prime=1.$ We are left with 6 parameters.
Since a solution is specified by fixing 5 parameters, we are left with one parameter, which we can choose freely. We will choose this free parameter to be the first horizon size $R_1$. This freedom is equivalent to specifying the mass $M_0$ in the SdS solution in section \ref{subSdS}. Since there is no asymptotically flat region in our solutions, we prefer not to talk about a mass; meanwhile the radius of the first horizon is clearly a physical quantity (one could obtain it by measuring the area $4\pi R_1^2$ of the horizon).

In practical calculations we thus first fix the first horizon size $R_1$. We then solve the field equations starting from some small $\epsilon$ away (typically of order $10^{-5}$) from the two horizons, with initial conditions given by the expansions in Eqs.~(\ref{eq:phiTaylorBHgg}, \ref{eq:fTaylorBHgg}, \ref{eq:RTaylorBHgg}), in the direction of positive $f$. As explained above, at the first horizon $r_1=0, f^\prime_1=1$ and thus we only have a single free parameter there, namely $\varphi_1.$ From both horizons we solve the equations of motion until $f(r)$ reaches a maximum. At this location we match the two half-solutions together. This has two consequences: the first is that $f^\prime$ is automatically continuous, and the second is that the coordinate location $r_2$ of the second horizon is implied. We then use a Newtonian algorithm to optimise all of the free parameters, until the fields and their derivatives are continuous at the matching location with sufficient precision. Experience has shown that it is necessary to optimise the free parameters to 3 significant digits in order to obtain solutions that are smooth to a satisfactory degree, although in principle the procedure may of course be continued to any accuracy desired.

\begin{figure}[ht!]
	\centering
	\includegraphics[width=0.5\textwidth]{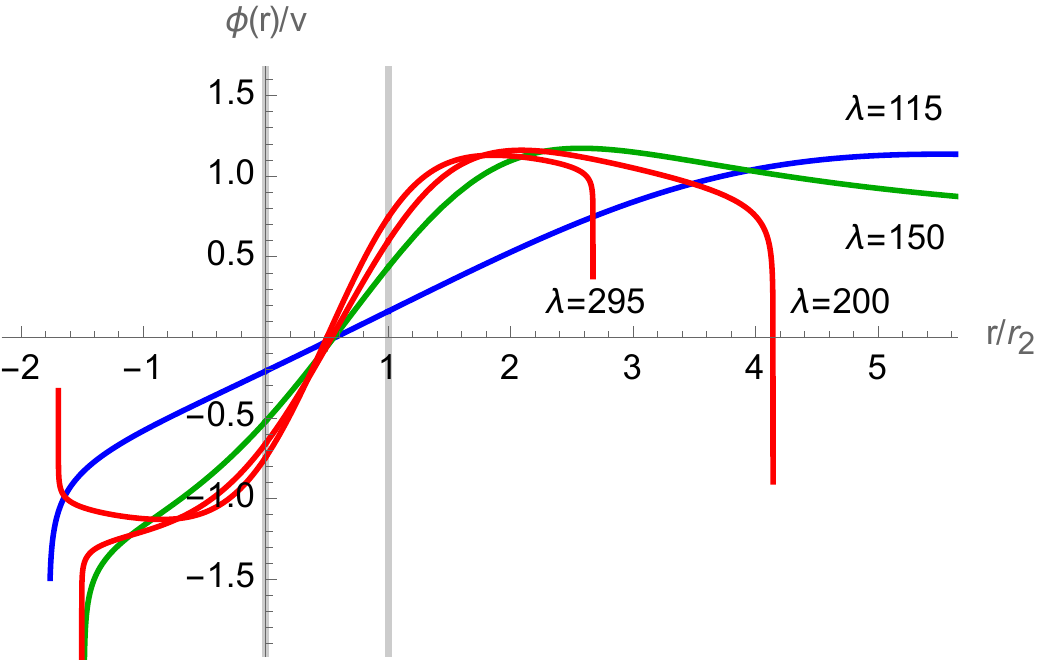} \\ \includegraphics[width=0.5\textwidth]{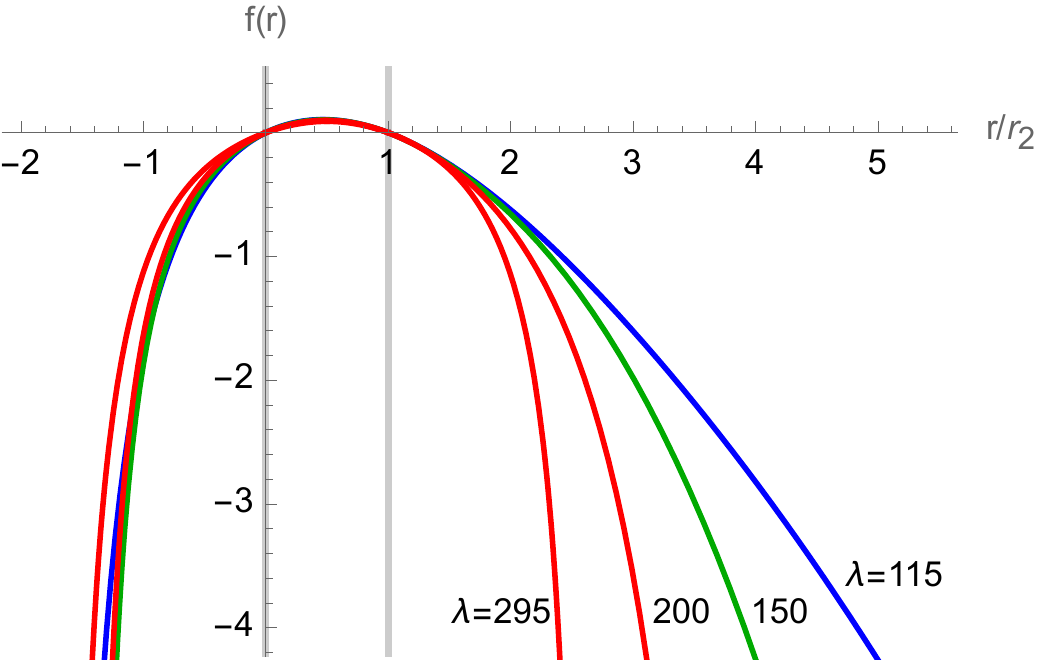} \\ \includegraphics[width=0.5\textwidth]{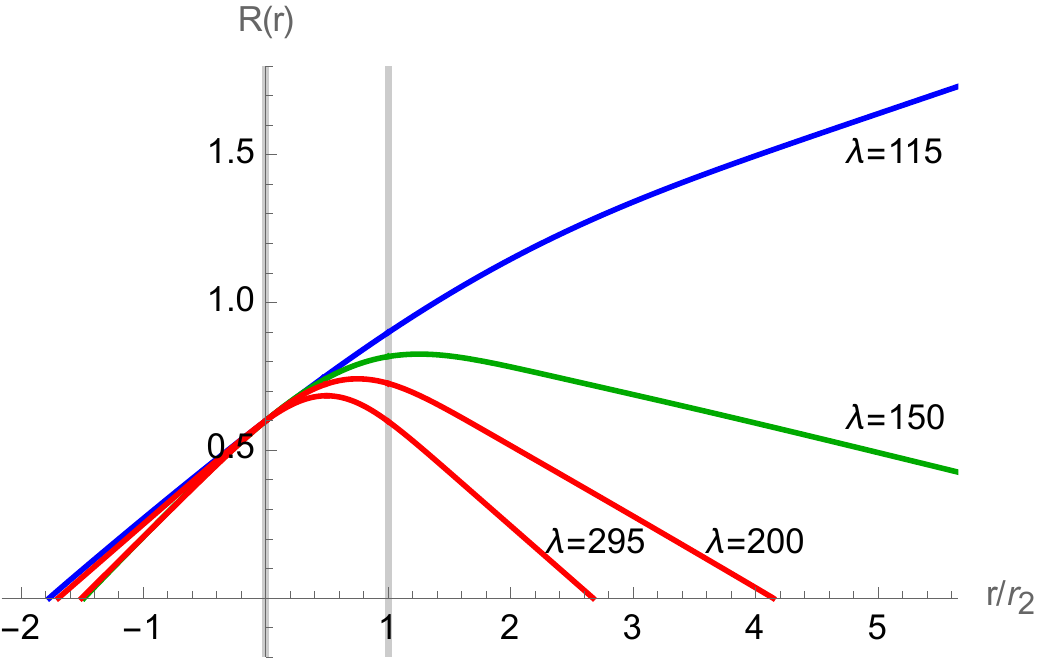}
	\caption{Examples of solutions with the potential ``width'' fixed at $v=0.18$ for various coupling constants $\lambda.$ For all of these solutions the first horizon size is set at $R_{1}=0.6$ and resides at $r=0.$ The location of the horizons is indicated by the thick vertical grey lines (one being at the origin $r=0$). The scalar field is rescaled such that the minima of the potential reside at $\pm 1.$ Also, these solutions have been plotted as functions of $r/(r_2-r_1)=r/r_2$ in order to facilitate a comparison; the second horizon then always resides at $r/r_2=1.$ See the main text for a full description.}
	\label{fig:examples}
\end{figure}

\begin{table}
\begin{tabular}{ |p{3cm}|p{3cm}|p{3cm}|p{3cm}|p{3cm}|}
		\hline
		$\lambda$ & \textbf{$\varphi_1$}& \textbf{$\varphi_2$} & \textbf{$f'_2$} & \textbf{$R_{2}$} \\
		\hline
		115 & -0.0376 & 0.0291 & 0.765 & 0.898 \\
		\hline
		150 & -0.0935 & 0.0784 & 0.828 & 0.817 \\
		\hline
		200 & -0.118 & 0.107 & 0.900 & 0.726 \\
		\hline
		295 & -0.134 & 0.134 & 1.000 & 0.598 \\
		\hline
		\end{tabular}
\caption{Optimised horizon values for the solutions in Fig. \ref{fig:examples}, given here to three significant digits.}\label{table1}
\end{table}

It turns out that the most interesting solutions can be found by looking for scalar field profiles that interpolate across the potential barrier at $\varphi=0.$ We will mainly concentrate on solutions that interpolate once across the barrier (in later sections we will also comment on solutions with more interpolations). We found solutions of three types -- see Fig. \ref{fig:examples} for an illustration, and table \ref{table1} for the corresponding optimised parameter values.

First, there are solutions in which $R(r)$ grows monotonically. We depict them in blue. These solutions are rather close analogues of the Schwarzschild-de Sitter solution. The smaller horizon is a trapped surface enclosing a singularity where $R(r)$ reaches zero size, in other words it is a black hole. In the approach to the singularity the solution is of approximate Kasner form, with $f$ and $\varphi$ both diverging \cite{Lavrelashvili:2021rxw}. The second horizon is a cosmological horizon, beyond which the scalar field undergoes damped oscillations around one of the potential minima. Asymptotically the solution becomes the de Sitter solution, with $f$ negative and growing in magnitude as $r^2$ and $R(r)$ growing linearly, while $\varphi$ sits in one of the potential minima and provides the required vacuum energy. Solutions of this form were discovered in \cite{Torii:1999uv}, where it was realised that these solutions demonstrate that non-trivial scalar field configurations may exist in and around black holes in theories with suitable potential maxima. 
This would be of interest with regard to no-hair arguments in these theories,  
but however it was also suggested that such solutions are unstable. 
``Blue'' solutions were the only ones known to date. 
We have discovered two new types of solutions, which differ more considerably from SdS.

In the second type of solutions $R(r)$ grows monotonically in between the two horizons, but nevertheless reaches zero beyond both horizons. We depict these cases in green. The smaller horizon again represents a black hole, and behaves identically to the blue solutions described above. The second horizon is a cosmological horizon, but it now encloses a re-collapsing universe. In other words, beyond the second horizon the scalar field evolution is so significant that the radii of spatial spheres start decreasing again beyond a certain location where a maximum radius is attained. The crunching universe then behaves very much like the interior of a black hole, with both $f$ and $\varphi$ diverging as $R$ approaches a second zero. However, this crunching universe is shielded from view to any observer residing in between the two horizons. In our numerical examples, we found that quite often (though not always) the maximum of $R(r)$ is reached very close beyond the second horizon, and the example in Fig. \ref{fig:examples} fits this characterisation.

Finally, in the third type of solution, $R(r)$ already has a maximum in between the two horizons. These solutions are shown in red. Here the scalar lump (implied by the strong scalar field profile) curves the space to such an extent that $R(r)$ reaches a maximum size and decreases again before the second horizon is reached. Now there is very little difference between the two horizons, only their sizes are in general different. Both are black hole horizons (they are trapped surfaces), beyond which $f$ and $\varphi$ diverge at a finite ``distance''. The scalar lump is thus surrounded by black holes in \emph{both} radial directions. Hypothetically travelling from one black hole towards the scalar lump one would find oneself in an increasingly large space, only to find the space to be shrinking again towards a second black hole horizon as one continues in the same direction beyond the scalar lump. These red solutions arise at larger values of the coupling constant $\lambda,$ which makes sense as the larger self-interaction of the scalar allows for higher concentrations of scalar field gradient energy, i.e. ``heavier'' lumps. The red solution at $\lambda=295$ shows a case where the solution is completely symmetric to good accuracy. This should be contrasted with the Nariai solution, in which the two horizons are also of equal size but the spacetime is in fact non-singular and the sphere is of constant size. With the interpolating scalar field this is thus morphed into a double black hole solution.

\begin{figure}[ht!]
	\centering
	\includegraphics[width=0.4\textwidth]{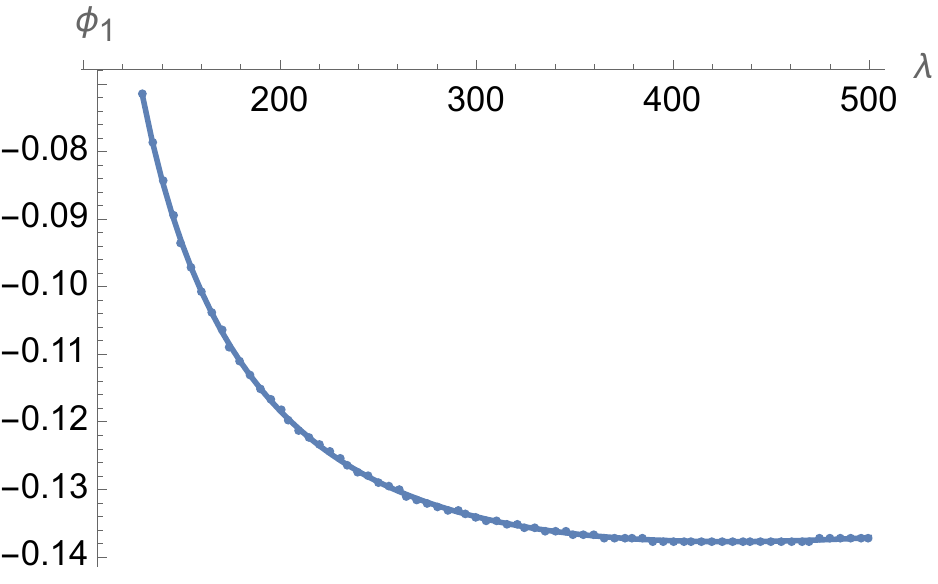} \quad \includegraphics[width=0.4\textwidth]{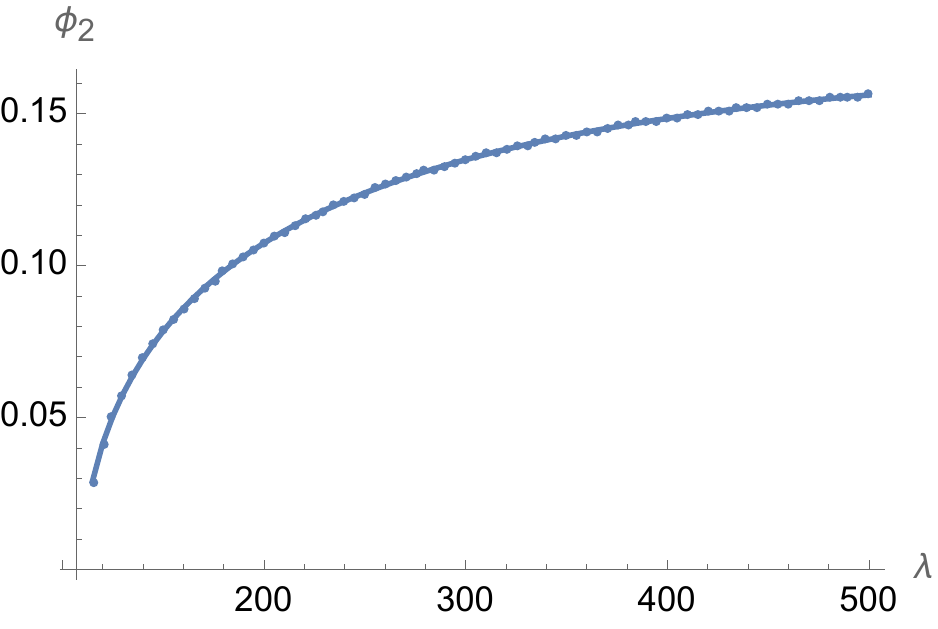}\\
	\includegraphics[width=0.4\textwidth]{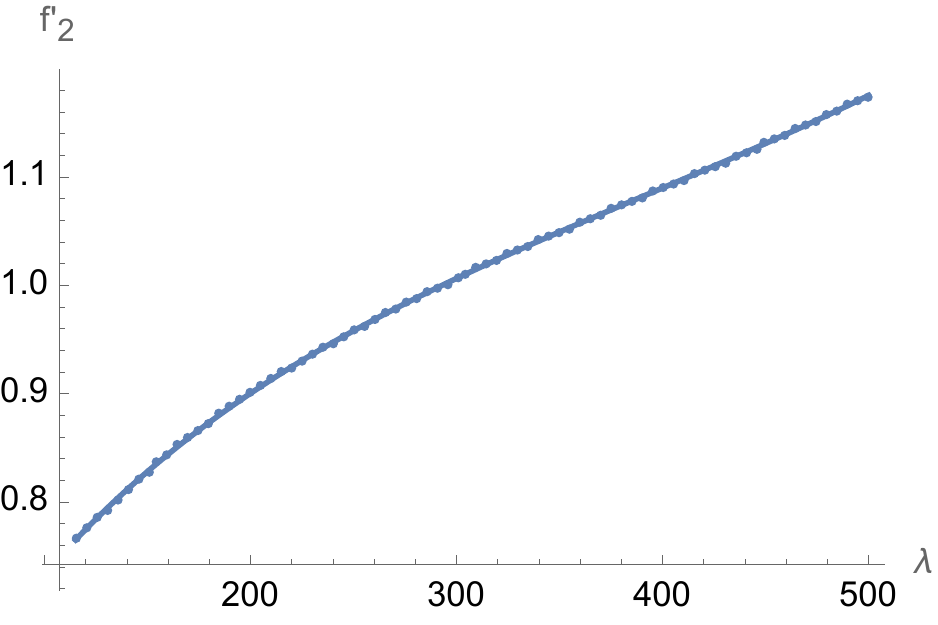} \quad \includegraphics[width=0.4\textwidth]{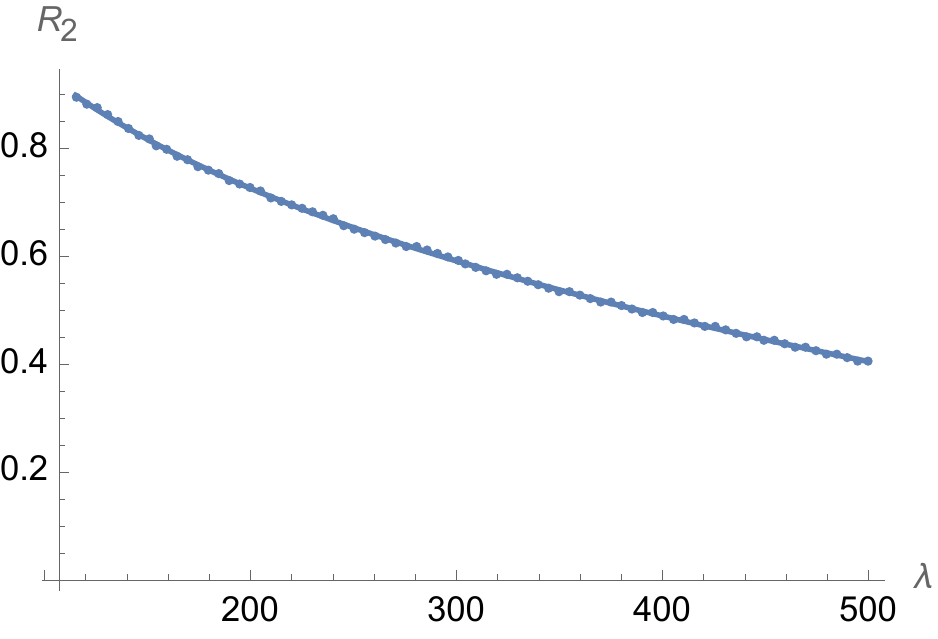}
	\caption{Optimised values of the horizon parameters, for fixed potential width $v=0.18$ and increasing coupling constant $\lambda.$ For these solutions one horizon is fixed at size $R_1=0.6$.}
	\label{fig:optimised}
\end{figure}

The examples above already show that the existence of solutions, and of the type of solution, depends crucially on the parameters of the potential. It is generally the case that at large coupling constant $\lambda$ solutions continue to exist. This can be inferred for instance from series of solutions with increasing coupling, see Fig. \ref{fig:optimised}. One can see that with increasing $\lambda,$ the optimised parameters vary less and less, and there is no indication of an obstruction to the existence of solutions. However, the existence of solutions depends quite strongly on the potential width $v,$ and moreover the character of the solutions (i.e. blue, green or red) depends crucially on the $v,\lambda$ combination. Furthermore, all properties depend on the value of the overall free parameter $R_1,$ i.e. on the size of a ``seed'' horizon.

\begin{figure}[ht!]
	\centering
	\includegraphics[width=0.5\textwidth]{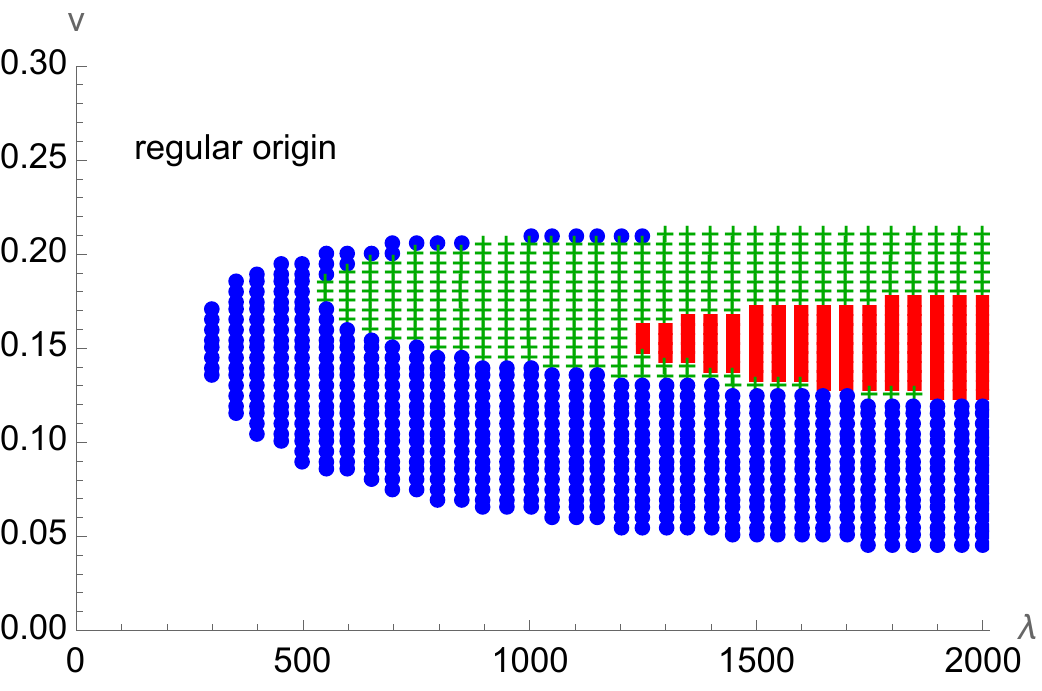} \\ \includegraphics[width=0.5\textwidth]{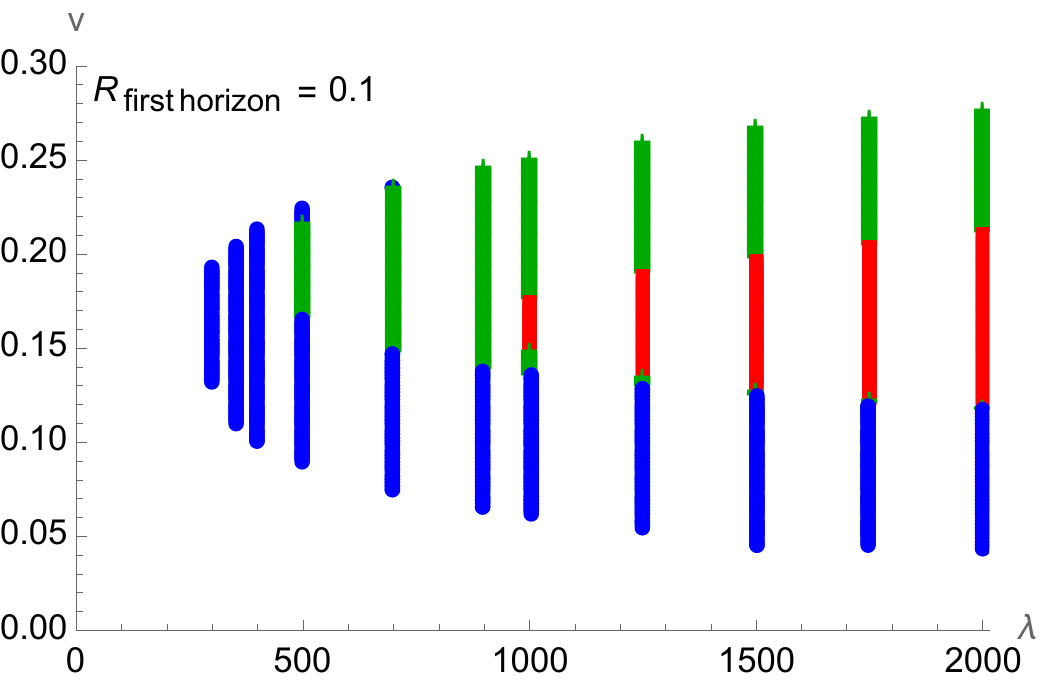}\\
	\includegraphics[width=0.5\textwidth]{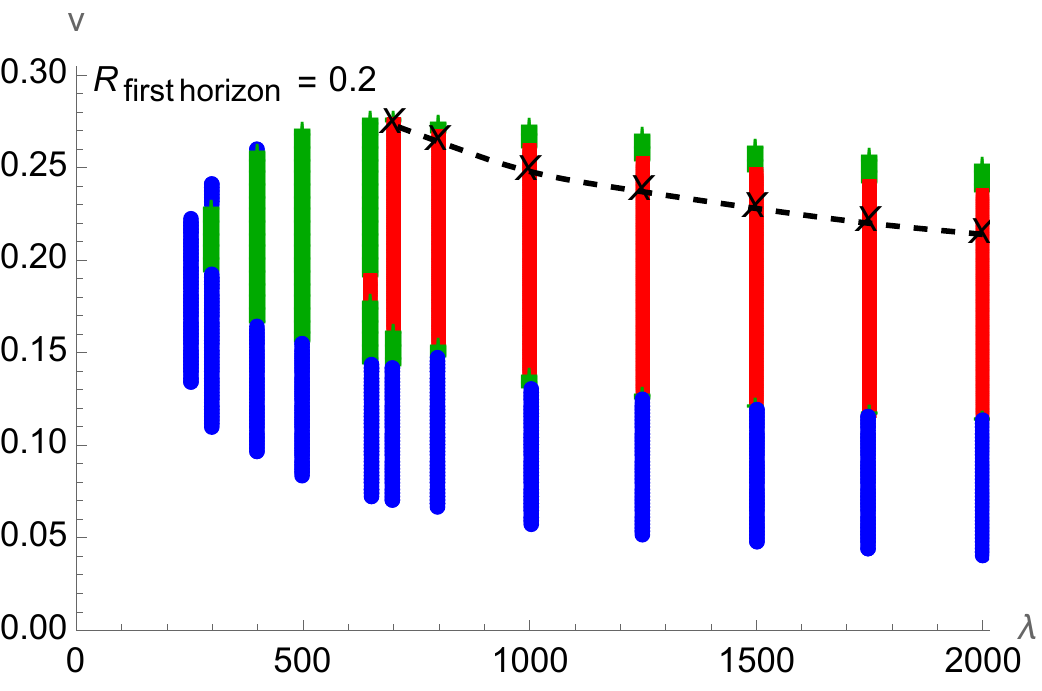} \\ \includegraphics[width=0.5\textwidth]{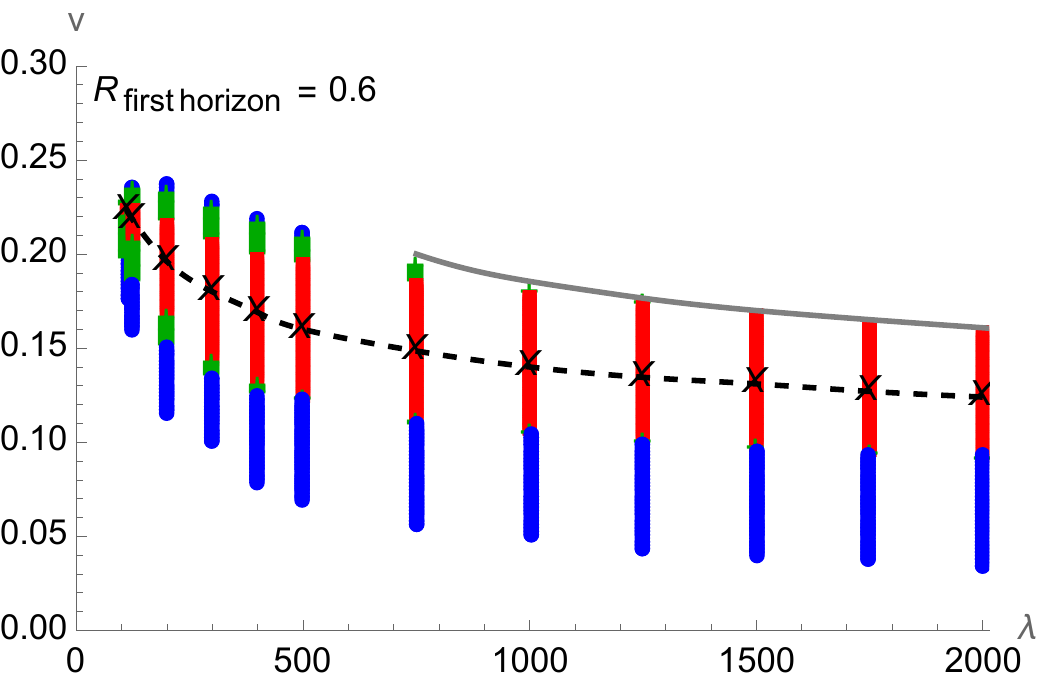}
	\caption{Overview plots for different sizes $R_1$ of the first horizon, as a function of the parameters $\lambda, v$ in the potential. We also included a reference plot at the top of the existence of solutions with a regular origin and a single horizon. Blue dots indicate the existence of solutions with monotonic $R(r),$ green crosses indicate solutions in which $R(r)$ turns around outside of a horizon, and red squares indicate double-black hole solutions in which $R(r)$ turns around in between the two horizons. A full description is given in the main text.}
	\label{fig:overview}
\end{figure}

We have undertaken an extensive numerical survey of the existence and type of solutions for different values of the first horizon size $R_1.$ These are shown in Fig. \ref{fig:overview} for $R_1=0.1, \, 0.2,\, 0.6$ and for reference we have also included a graph of the existence of solutions with a single horizon, where the second horizon is replaced by a regular origin (this graph is reproduced from our earlier Ref. \cite{Lavrelashvili:2021rxw}). As is apparent, we have optimised series of solutions at closely spaced intervals in $v,$ but with larger gaps in $\lambda;$ this was done simply for numerical convenience. We have adapted the spacings in $\lambda$ to the case at hand, in order to reveal changing properties of the solutions. In the figures, we have marked each solution that we found by a symbol (blue dots, green crosses and red squares). Black x's denote symmetric solutions, and the dashed black lines link these for the cases where they exist. Note that some solutions may appear in two separate graphs, since the second horizon of a solution may happen to be of the same size as the first horizon of another solution. Rather than being an annoyance, we have used this property as a cross check of our results.

Several broad features can immediately be identified: compared to single horizon solutions, at small coupling $\lambda$ solutions with two horizons exist for larger values of the width $v$ of the potential. For large first horizon size $R_1$ they also exist for smaller $\lambda.$ In all cases red solutions are nested within green regions, which are themselves nested within blue regions, unless solutions cease to exist altogether. Although our graphs only show the parameter space up to $\lambda=2000,$ we have checked for various values of $v$ that solutions continue to exist up to $\lambda=10000,$ and in fact, as argued above, we see no obstruction to solutions existing at arbitrarily large $\lambda.$ However, at small $\lambda$ there are clear limits to the existence of solutions, and only for a limited range of the width $v$ solutions can be found.

\begin{figure}[ht!]
	\centering
	\includegraphics[width=0.3\textwidth]{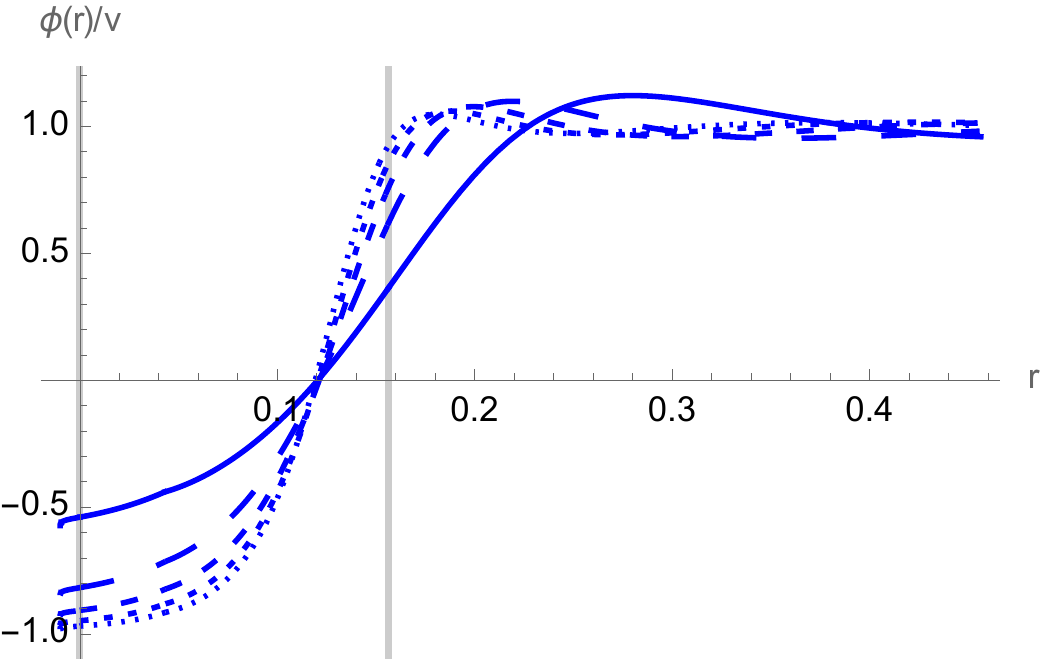} \,\, \includegraphics[width=0.3\textwidth]{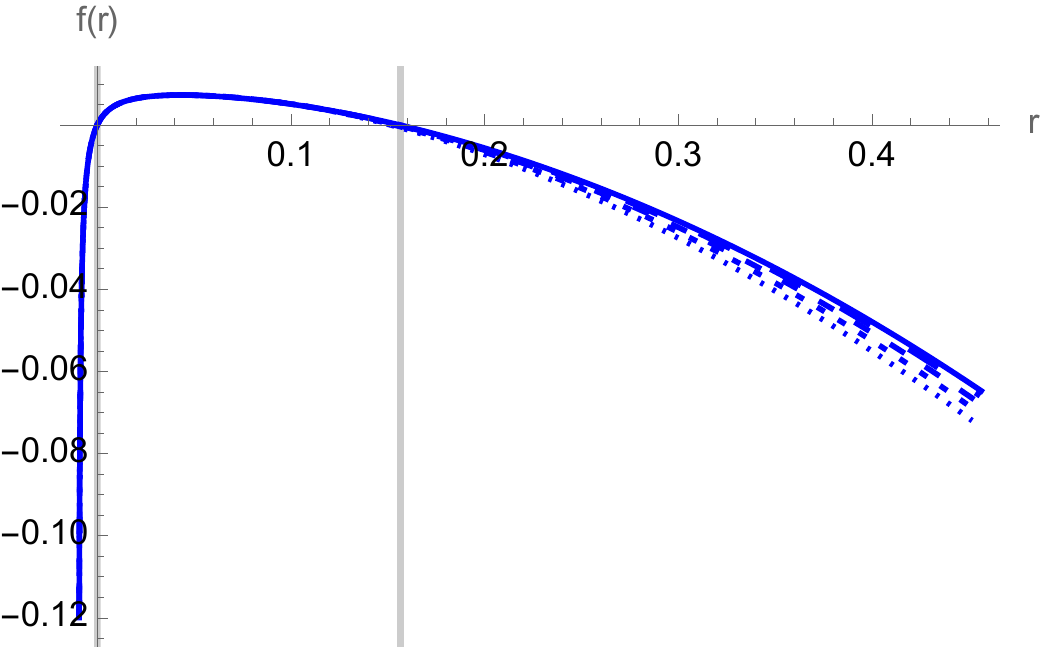} \,\, \includegraphics[width=0.3\textwidth]{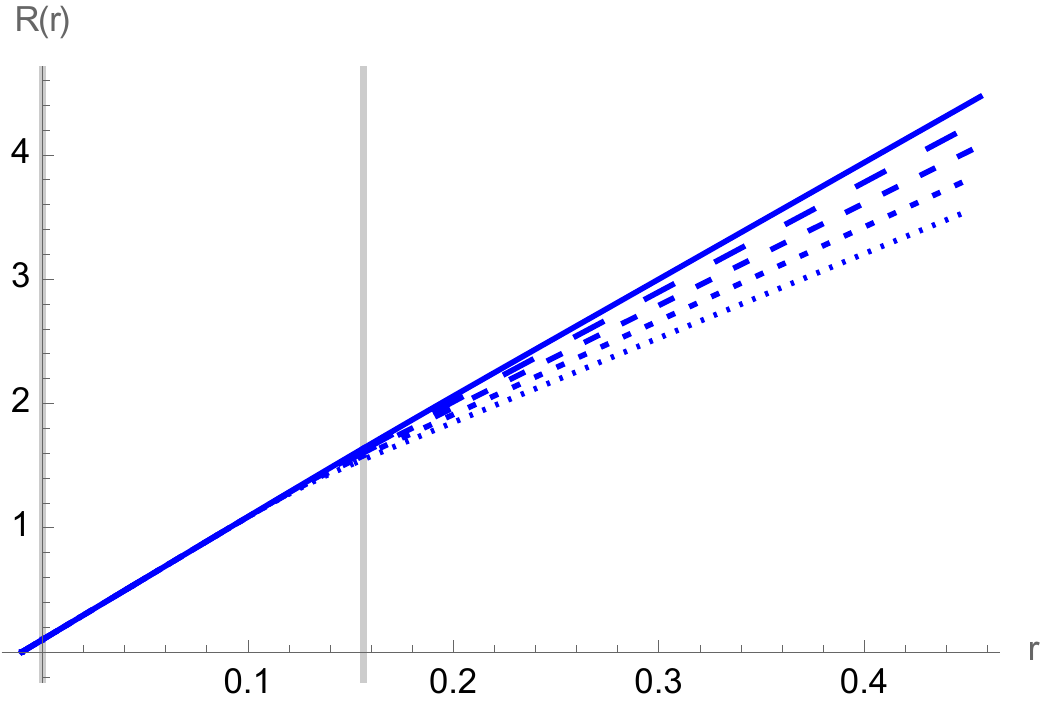}
	\caption{Series at $\lambda=2000$ and for small $v,$ with one horizon of size $R_1=0.1,$ near the boundary where solutions cease to exist. The values for the width of the potential are $v=0.068,0.062,0.056,0.050,0.044,$ smaller $v$ corresponding to the less dashed lines. One can see that as the limit is approached the solutions have ever smaller scalar field excursions inside the two horizons, and correspondingly $R(r)$ becomes ever straighter. The limiting solution (not shown here) is the excited SdS solution with the scalar sitting on top of the potential barrier.}
	\label{fig:nearboundary}
\end{figure}

When moving around in parameter space, solutions may cease to exist for several different reasons.\footnote{In our analysis of the limits of existence of solutions we disagree with the analysis performed in \cite{Torii:1998ir}.} It is easiest to look at examples to recognise the obstruction in each case. At small $v,$ i.e. at the bottom of the diagrams, solutions are always blue; a plot of several such solutions near the limit of existence are shown in Fig. \ref{fig:nearboundary}. Curves that are less prominently dashed, and thus more solid, are located at smaller $v,$ closer to the boundary. As the boundary is approached, the scalar field interpolates less and less far across the top of the potential barrier. Thus one can infer that the limiting solution is the unstable SdS solution with the scalar field sitting on top of the barrier.

\begin{figure}[ht!]
	\centering
	\includegraphics[width=0.3\textwidth]{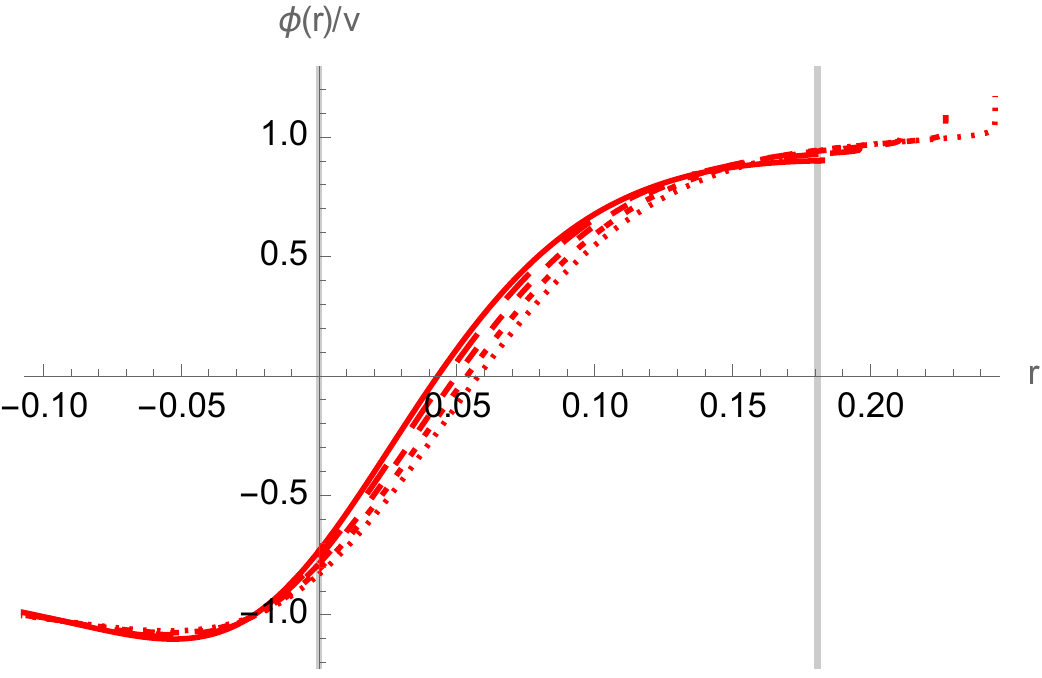} \,\, \includegraphics[width=0.3\textwidth]{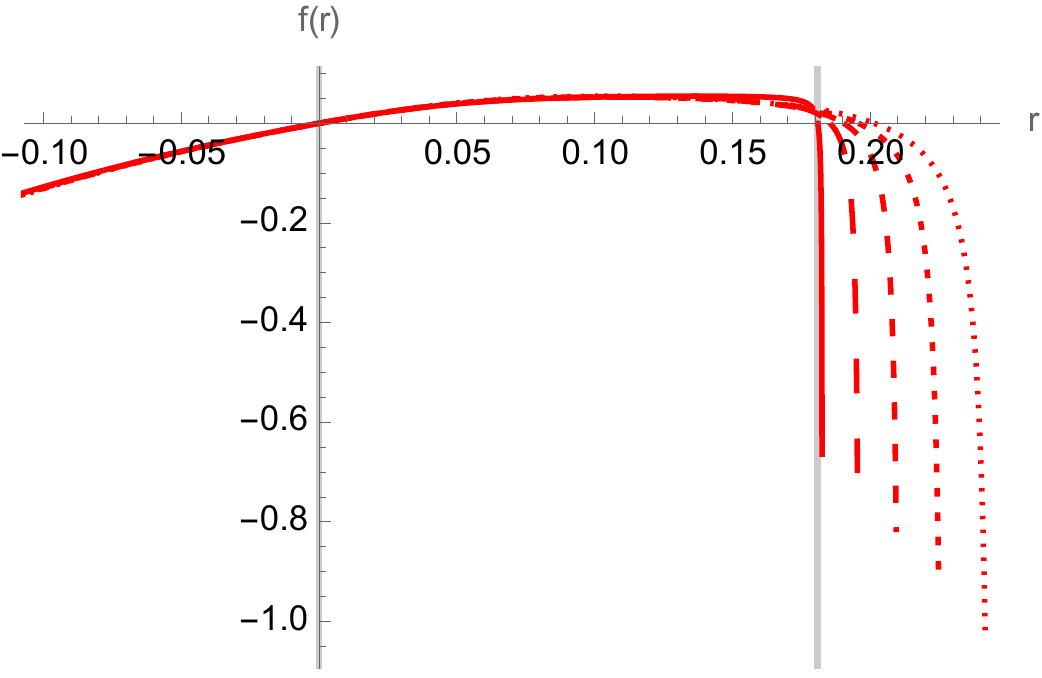} \,\, \includegraphics[width=0.3\textwidth]{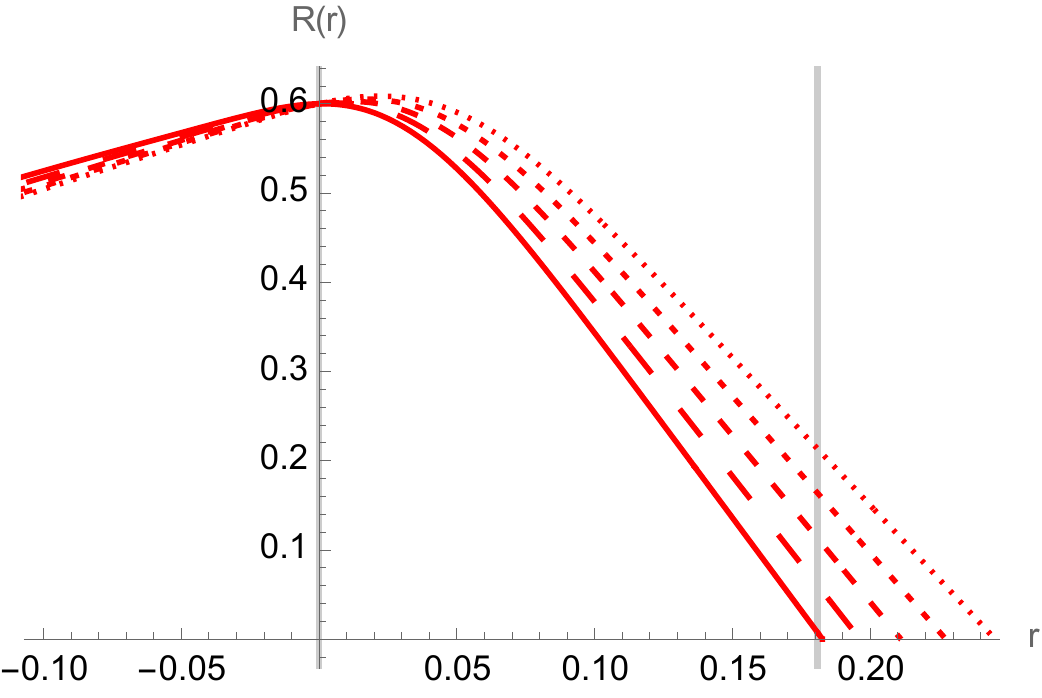}
	\caption{Series at $\lambda=1500$ and for large $v,$ with one horizon of size $R_1=0.6,$ near the boundary where solutions cease to exist. The limiting solutions have ever smaller second horizon radius, and disappear when this radius reaches zero size. Shown here are solutions with $\lambda=0.162,0.164,0.166,0.168,0.170$ with the larger $v$ values corresponding to curves that are progressively less finely dashed. Note that for these borderline solutions, the scalar field excursion remains significant. The limiting solution is not the excited SdS solution, as it was for small $v$ values, but rather a solution with one black hole reaching zero mass. This is however not a standard SdS solution, as $R$ still turns around in between the two horizons.}
	\label{fig:nearboundaryred}
\end{figure}

For sufficiently large $\lambda$ and large $v,$ the limiting behaviour is shown in Fig. \ref{fig:nearboundaryred}. Again, the boundary is reached for the curves that are more solid. Here one can see that the scalar field keeps interpolating between values that are close to the potential minima. Rather, this time it is the second horizon size $R_2$ that keeps shrinking, until it reaches zero size. The second horizon disappears. The limiting solution then has a single horizon and a regular origin at the other side. In other words, the limiting solution is one of the solutions in the top graph of Fig. \ref{fig:overview}, with the appropriate horizon size. We have indicated the location of the corresponding limiting solutions by the grey line in the bottom graph in Fig. \ref{fig:overview}. It is noteworthy that the limiting solutions are of different physical character also for different initial horizon sizes $R_1.$

\begin{figure}[ht!]
	\centering
	\includegraphics[width=0.45\textwidth]{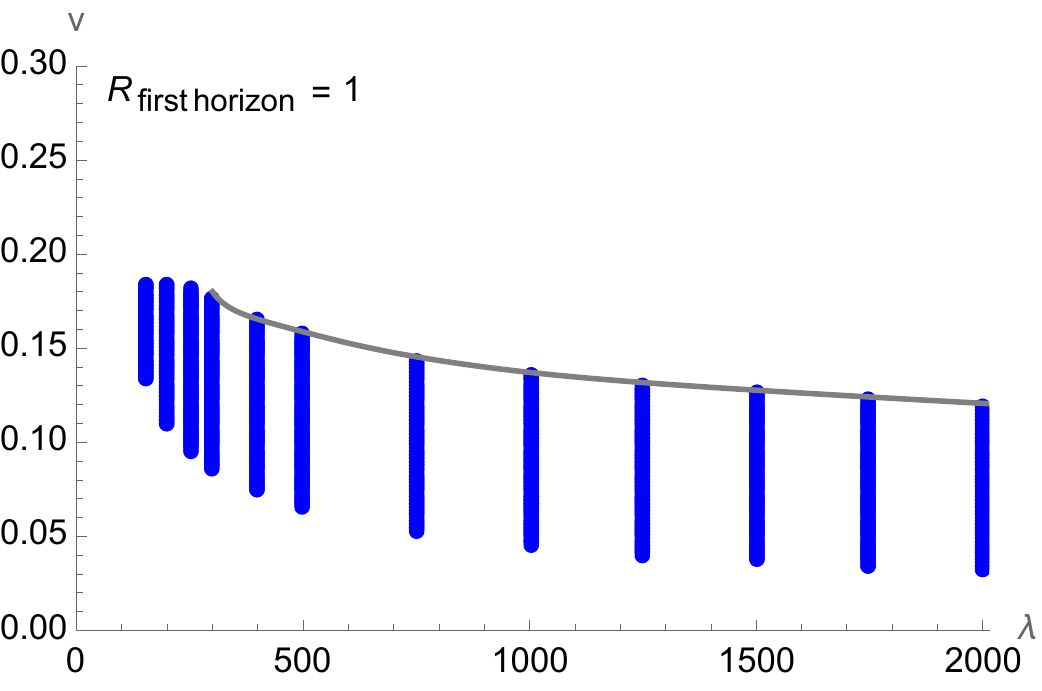}
	\caption{Overview plot of solutions with one horizon size fixed to the value $R_1=1.$ The grey curve corresponds to solutions with a regular origin and a single horizon.}
	\label{fig:summarylargeR}
\end{figure}

\begin{figure}[ht!]
	\centering
	\includegraphics[width=0.3\textwidth]{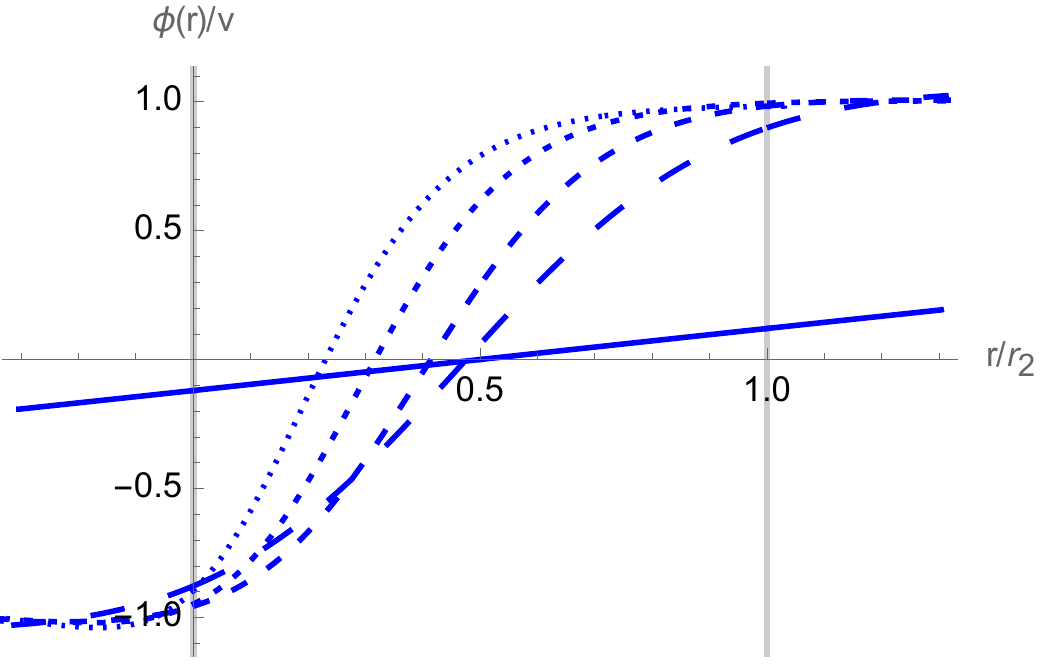} \,\, \includegraphics[width=0.3\textwidth]{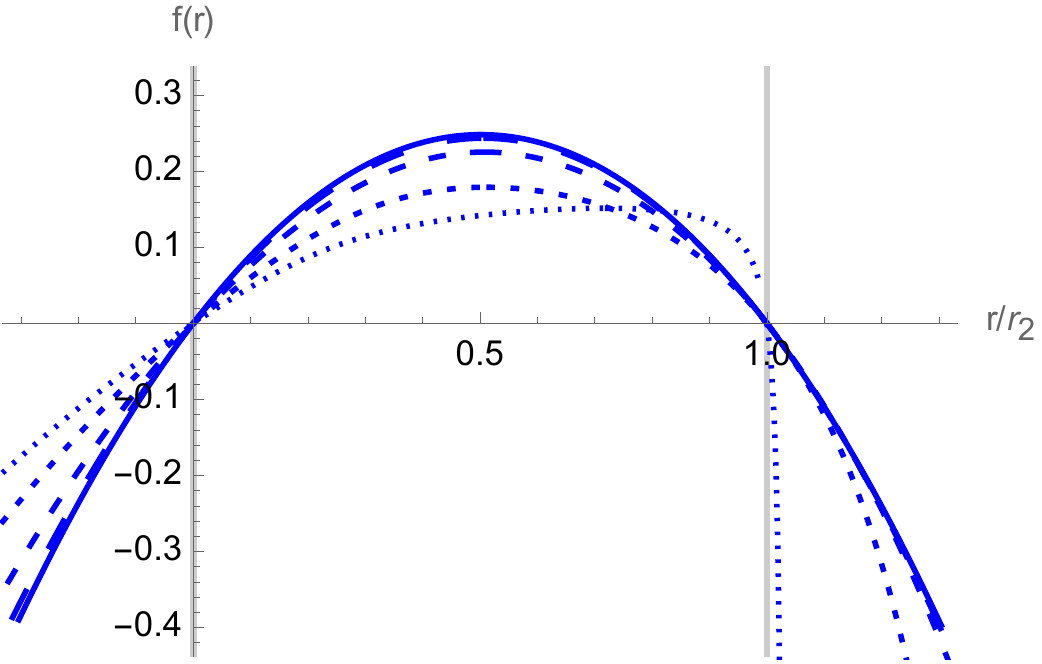} \,\, \includegraphics[width=0.3\textwidth]{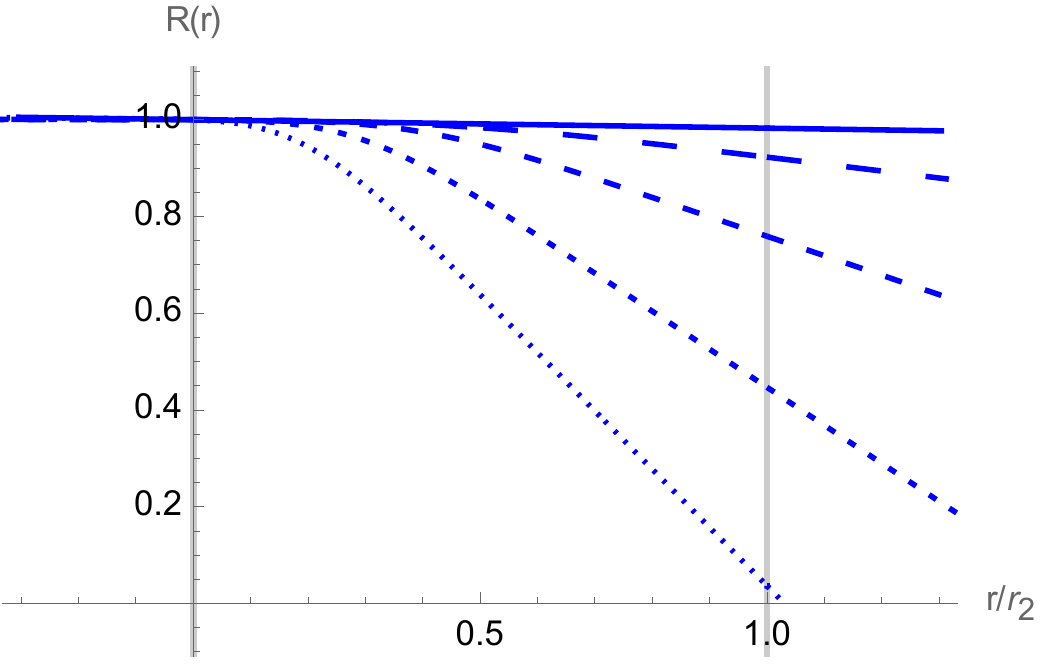}
	\caption{Series at $\lambda=1500$ and for one horizon of size $R_1=1.$ These solutions are sample solutions along a complete vertical slice through Fig. \ref{fig:summarylargeR}, with potential minima at $v=0.037,0.057,0.082,0.107,0.127$ this time with the larger $v$ values corresponding to curves that are progressively more finely dashed.  At small $v$ the limiting solution can be seen to be the Nariai solution, with constant $R$ and two horizons of equal size. At large $v$ the limiting solution is one where the horizon shrinks to zero size, i.e. the limiting solution will be one with just one horizon and a regular ending on the other side. Note that these solutions have been plotted as functions of $r/(r_2-r_1)=r/r_2$ in order to facilitate a comparison; the second horizon then always resides at $r/r_2=1.$}
	\label{fig:R1examples}
\end{figure}

Since single horizon solutions appear to be more important as limits of existence when $R_1$ is larger, we are prompted to also investigate the case where the first horizon size is very large. The case $R_1=1$ is shown in Fig. \ref{fig:summarylargeR}. When $R_1 \geq \frac{1}{\sqrt{\kappa V}},$ the Taylor expansion in Eq.~(\ref{eq:RTaylorBHgg}) implies that $R$ decreases towards the interior of the spacetime, which together with the fact that $R^{\prime\prime}$ is non-positive immediately implies that red solutions become impossible. And in fact for $R_1=1$ we only found blue solutions. Examples of solutions at fixed coupling $\lambda=1500$ and at representative values of $v$ are shown in Fig. \ref{fig:R1examples}. One can see that they interpolate between the two limiting behaviours identified thus far: an excited SdS solution on one side and a single horizon solution on the other.

\begin{figure}[ht!]
	\centering
	\includegraphics[width=0.45\textwidth]{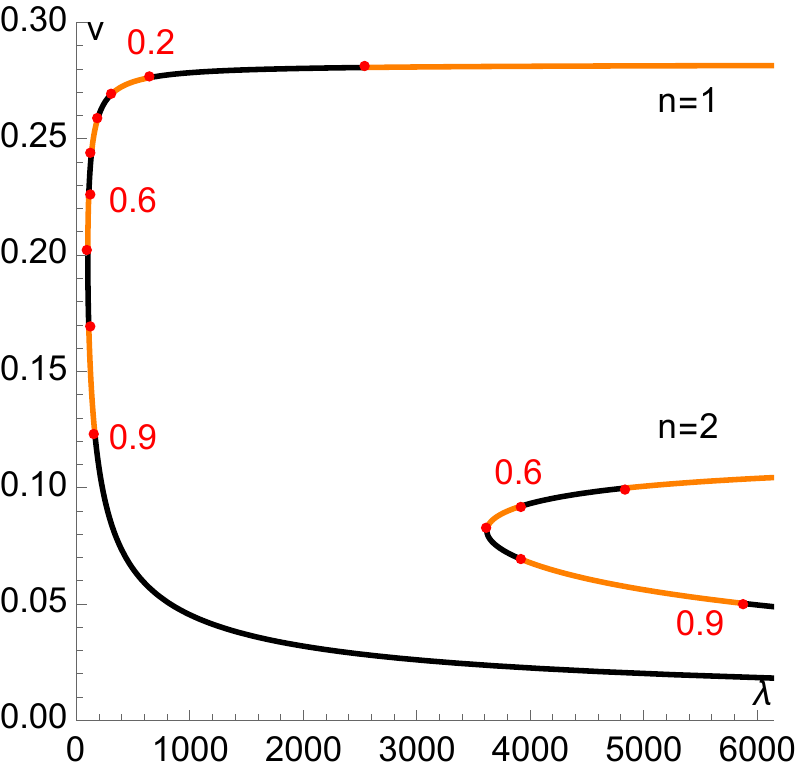}
	\caption{Limits of existence of solutions with two horizons of equal size, for various horizon sizes (selected values are shown in red) and with both a single ($n=1$) and two ($n=2$) interpolations. These points coincide with the limits of the symmetric solutions shown in Fig. \ref{fig:overview}, for the cases where symmetric solutions exist, i.e. for the plots with $R_1=0.2,\, 0.6$. The alternating colours are used to improve readability of the graph.}
	\label{fig:NariaiLimits}
\end{figure}

Continuing with our discussion of the limits of existence of solutions, we note that there is a third type of limiting behaviour: it concerns the fully symmetric solutions, when they exist. They can be seen as the black x's for $R_1=0.2$ and $R_1=0.6.$ Moving to smaller $\lambda,$ these solutions exhibit a smaller scalar field excursion across the potential barrier, much like the case shown in Fig. \ref{fig:nearboundary}. However, for these solutions the two horizon sizes remain equal, and in fact $R(r)$ becomes ever flatter due to the less prominent scalar field gradients. The limiting solution in this case is thus the Nariai solution, with the scalar field sitting on top of the potential barrier. Interestingly, in this case the limiting points can be found analytically, similarly to perturbations of single horizon solutions \cite{Torii:1999uv}. The (excited) Nariai solution is given by
\begin{align}
R=\frac{1}{\sqrt{\kappa V}} = \frac{1}{\sqrt{\Lambda_{eff}}}\,, \qquad f(r)=1-\Lambda_{eff}r^2\,, \label{eq:Nariaitop}
\end{align}
where $\Lambda_{eff} = 1 + \frac{\kappa}{4}\lambda v^4$ is the effective (constant) vacuum energy at the top of the potential barrier. The scalar equation of motion \eqref{eq:phi} is solved trivially with $V_{,\varphi}=0.$ However, perturbations around this solution may be non-trivial, and can indicate to us the limits of existence of solutions with an interpolating scalar field. For this, we have to look at perturbations $\delta\varphi$ of the scalar field equation. Around the Nariai solution, we find
\begin{align}
& \delta\varphi^{\prime\prime}+\frac{f^\prime}{f}\delta\varphi^\prime + \frac{V_{,\phi\phi}}{f}\delta\varphi = 0\,, \label{scalarpert} \\
\rightarrow \, & \left(f \, \delta\varphi^\prime \right)^\prime + \lambda v^2 \delta \varphi = 0\,.
\end{align}
If we rescale $z\equiv \sqrt{\Lambda_{eff}}r,$ the above equation turns into
\begin{align}
\left[ (1-z^2)\delta\varphi_{,z}\right]_{,z} + \frac{\lambda v^2}{1+\frac{\kappa}{4}\lambda v^4}\delta\varphi = 0\,,
\end{align}
which can be recognised as a Legendre equation. The solutions are the $\nu$th Legendre functions, with $\frac{\lambda v^2}{1+\frac{\kappa}{4}\lambda v^4} = \nu (\nu + 1).$ With the condition that at the first horizon the scalar should remain at the top of the barrier, i.e. that $\delta\varphi(0)=0,$ this restricts $\nu$ to be a positive odd integer. For $\nu=2n-1,$ this applies to solutions that interpolate $n$ times across the potential barrier. Taking into account the relation \eqref{eq:Nariaitop} between the horizon radius and the parameters of the potential, we end up with the critical values $v_c, \lambda_c$ that indicate the limit of existence of solutions that have equal horizon sizes $R_h<1$ and an interpolating scalar field,
\begin{align}
v_c = \frac{4}{\kappa}\frac{1-R_h^2}{\nu(\nu+1)}\,, \qquad \lambda_c = \frac{4}{\kappa V_c^4}\left( \frac{1}{R_h^2} - 1\right)\,. \label{eq:Nariailimit}
\end{align}
These points are shown numerically in Fig. \ref{fig:NariaiLimits}, and they can be verified to agree very precisely with the limits of the black dashed curves in the graphs for $R_1=0.2,\, 0.6$ in Fig. \ref{fig:overview}.

This concludes our description of the parameter space of existence of the double-horizon solutions.


\section{Geodesics}

In order to clarify the structure of the solutions we have descried in the previous section, we may also look at geodesic motions on these spacetimes. A geodesic curve $x^\mu(\lambda)=(t(\lambda),r(\lambda),\theta(\lambda),\phi(\lambda))$ with affine parameter $\lambda$ satisfies the condition of auto-parallel transport,
\begin{align}
u^\mu \nabla_\mu u^\nu = 0\,,
\end{align}
where we denoted $u^\mu = \frac{dx^\mu}{d\lambda}.$ The analysis is greatly simplified by symmetries of the spacetime, which manifest themselves by the existence of Killing vectors $k^\mu$ satisfying $\nabla_\mu k_\nu + \nabla _\nu k_\mu = 0.$ It follows immediately that $u^\mu \nabla_\mu (k_\nu u^\nu)=0,$ i.e. along the curve the quantity $k_\nu u^\nu$ is a constant of motion. Our spacetime is of the form
\begin{align}
ds^2 = - f(r)dt^2 + \frac{dr^2}{f(r)} + R^2(r) (d\theta^2 + \sin^2 (\theta) d\phi^2)\,.
\end{align}
We consider the geodesics to lie in the equatorial plane $\theta = \pi/2.$ Since the metric contains no dependence on $t$ nor $\phi$ we immediately obtain the two Killing vectors $(\partial_t)^\mu = (1,0,0,0)$ and $(\partial_\phi)^\mu = (0,0,0,1),$ leading to a conserved energy $E$ and angular momentum $L,$
\begin{align}
E = f(r) \dot{t}\,, \qquad L = R^2(r) \dot\phi\,.
\end{align}
For a timelike geodesic, we have
\begin{align}
-f \dot{t}^2 + \frac{\dot{r}^2}{f} + R^2 \dot\phi^2 = -1\,,
\end{align}
which can be rearranged as the equation of one-dimensional motion in an effective potential $U(r):$
\begin{align}
\frac{1}{2}\dot{r}^2 + U(r) = \frac{1}{2}E^2\,, \qquad U(r) = \frac{1}{2}f(r)\left( 1 + \frac{L^2}{R^2(r)}\right)\,.
\end{align}

\begin{figure}[ht!]
	\centering
	\includegraphics[width=0.3\textwidth]{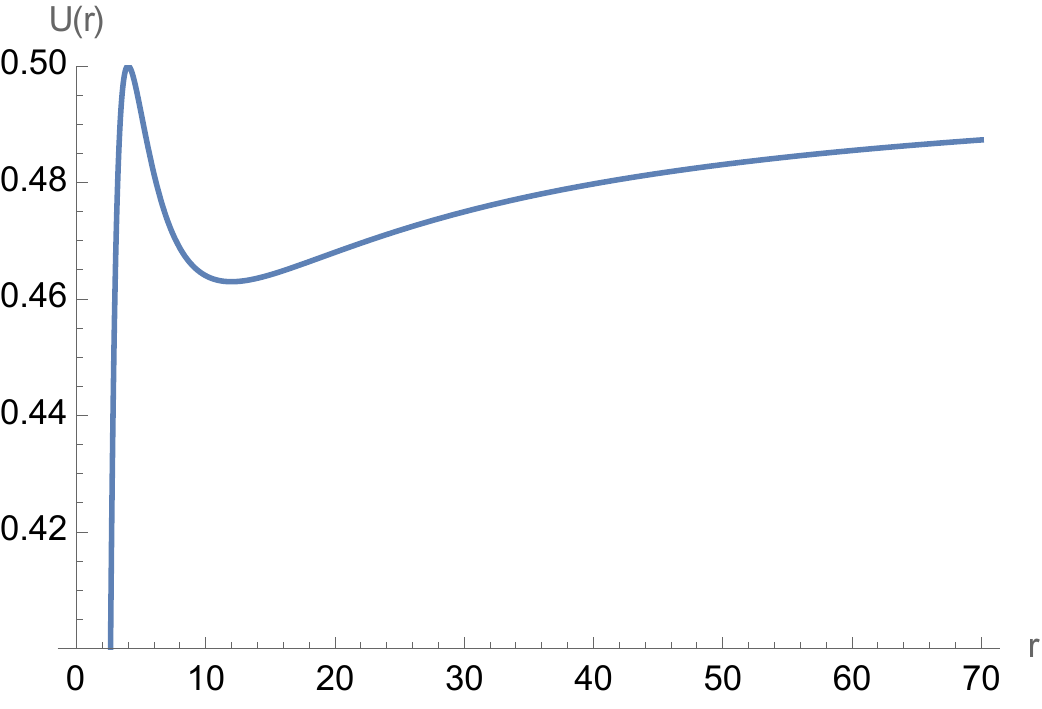} \includegraphics[width=0.3\textwidth]{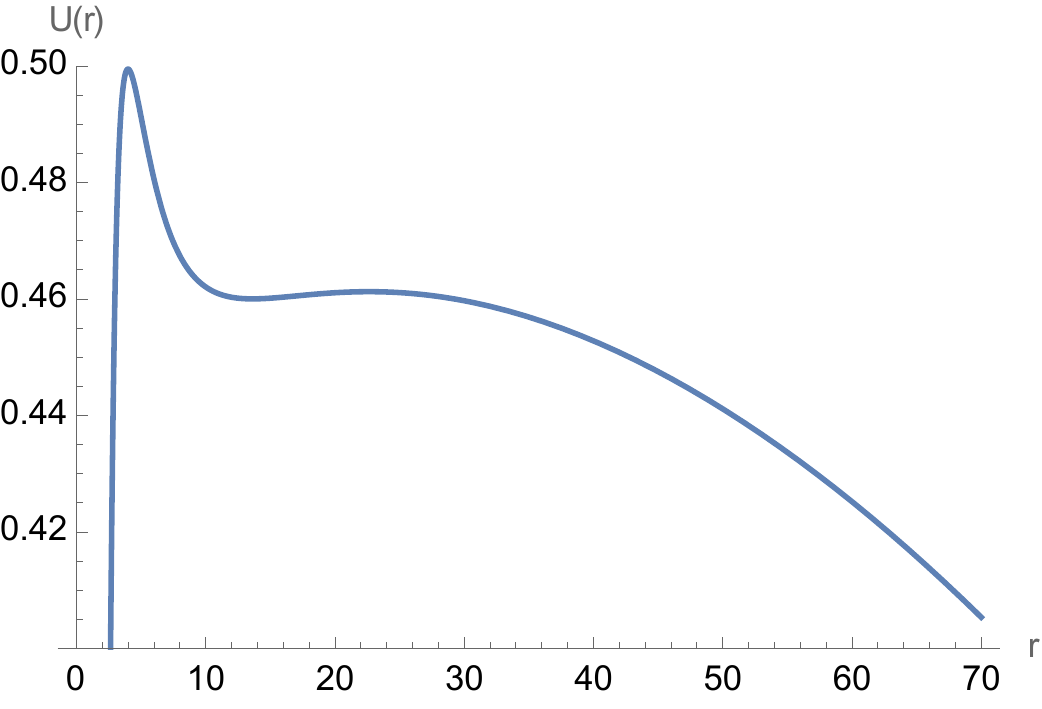}\includegraphics[width=0.3\textwidth]{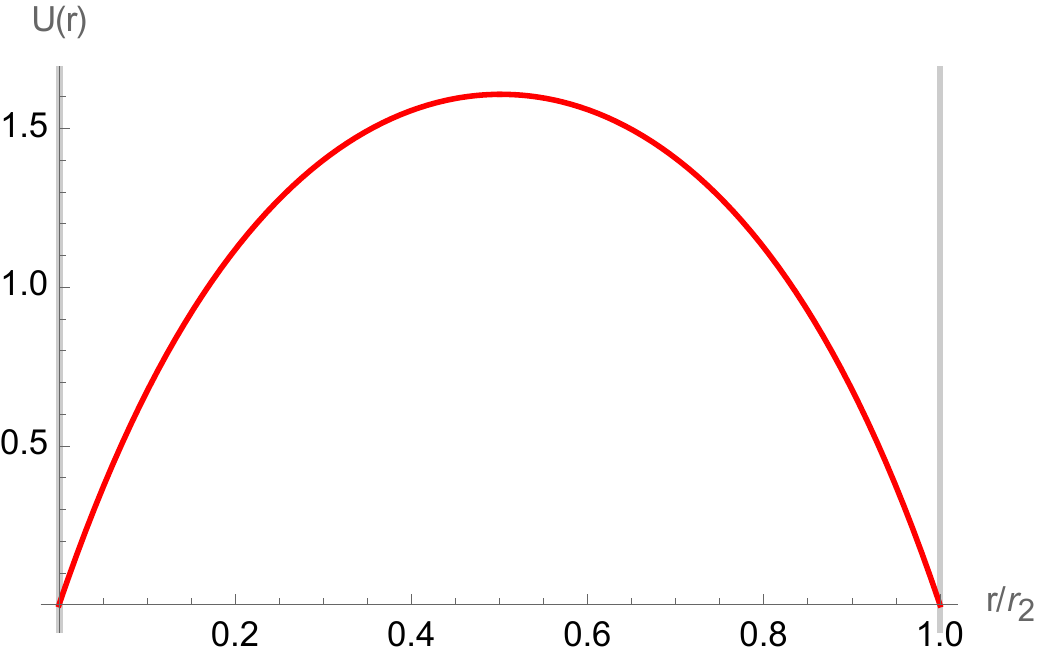}
	\caption{Effective potentials for radial motion. Left panel: Schwarzschild black hole (with mass set to unity and angular momentum $L=4$). Middle panel: Schwarzschild-de Sitter spacetime (with mass set to unity, angular momentum $L=4$ and cosmological constant $\Lambda=10^{-4}$). Right panel: scalar lump surrounded by two black holes (solution with $\lambda=295, v=0.18, R_1=0.6$).}
	\label{fig:effpotSdS}
\end{figure}

In the Schwarzschild geometry, we have $f(r)=1-\frac{2M}{r}$ and $R(r)=r.$ The effective potential, for intermediate values of angular momentum, is shown in the left panel of Fig.~\ref{fig:effpotSdS}.  Far away, any test particle is drawn towards the black hole, and the same occurs if the test particle approaches too closely to the black hole. There also exists a stable circular orbit. At large $r$ the potential approaches a constant value, since the attractive force of the black hole diminishes continuously. For Schwarzschild-de Sitter the potential is shown in the right panel, for the same value of the angular momentum and a small cosmological constant. Near the black hole, the situation is analogous, but now, far away from the black hole, the cosmological constant takes over and pushes any test particle away from the black hole. This is how, despite the metric being static, a test particle experiences the expansion of space sufficiently far from the black hole. The effective potential reaches minus infinity as $r\to \infty.$

\begin{figure}[ht!]
	\centering
	\includegraphics[width=0.4\textwidth]{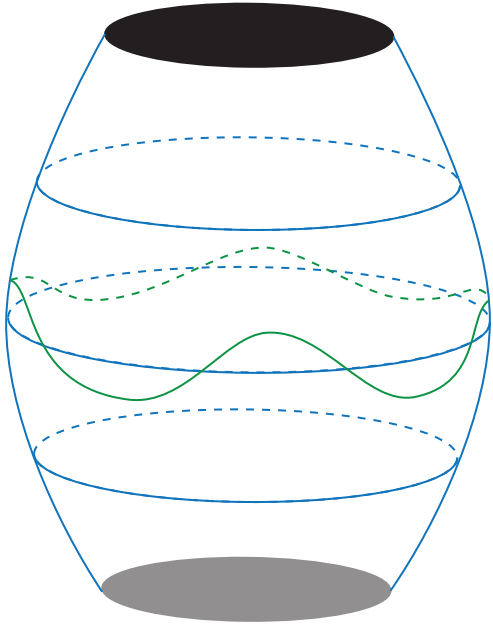}
	\caption{A cartoon of the solutions with non-monotonic $R(r).$ Depicted here are only the $r$ direction and a transverse circle instead of transverse spheres. The spacetime is capped off by two black holes, one on each Pole. In green we show a particle undergoing accelerated motion near the Equator, which itself represents an unstable orbit. If the particle travels too close to either black hole, it will inexorably be pulled in. The vacuum energy changes across the spacetime: it interpolates between the two sides of the scalar potential barrier as one traverses from one black hole to the other.}
	\label{fig:sketch}
\end{figure}

For the solutions with non-monotonic $R(r),$ the effective potential looks very different, and is shown in the right panel of Fig. \ref{fig:effpotSdS}, again for intermediate values of angular momentum.  In that case, one is either attracted towards the first or the second horizon, and only an unstable circular orbit exists in the middle. For strongly non-monotonic $R$ one might have imagined that it would be possible to have a stable region in the middle of the spacetime, with a stable circular orbit, because the attraction towards both horizons is overcome by a combination of the expansion of space and the angular momentum of the orbit. However, the field equations preclude this possibility, i.e. the field equations imply that $U(r)$ cannot develop a minimum. See Fig.~\ref{fig:sketch} for a sketch of the ``doomsday'' spacetimes with two spherically symmetric black holes.

\section{Black hole seeded vacuum decay}

One context in which the solutions we have discussed may find application is that of cosmological phase transitions. First order phase transitions in particular may have played an important role in the early universe, and have been studied for several decades now. At zero temperature they typically proceed as tunneling events from a local minimum (false vacuum) to the global minimum
(true vacuum) of the potential. While in flat spacetime the theoretical description of these events  is very well established
\cite{Coleman:1977py,Callan:1977pt,Coleman:1978ae,Coleman:1987rm}, the inclusion of gravity leads to Coleman-De Luccia (CdL) bounce solutions \cite{Coleman:1980aw}. These  introduce some open questions, like the negative mode problem \cite{Lavrelashvili:1985vn,Tanaka:1992zw,Lavrelashvili:1998dt,Khvedelidze:2000cp,Lavrelashvili:1999sr,Gratton:2000fj,Koehn:2015hga,Bramberger:2019mkv,Jinno:2020zzs}, that have not been fully resolved to date.
For flat potential barriers CdL bounces do not exist \cite{Hawking:1981fz,Jensen:1988zx,Balek:2004sd,Battarra:2013rba}. In this situation the phase transition is governed by the Hawking-Moss (HM) instanton \cite{Hawking:1981fz},
which exists for any scalar field potential with a local maximum. In the modern interpretation \cite{Brown:2007sd} the HM instanton describes a thermal fluctuation within a horizon volume at the temperature of de Sitter space.
In addition to CdL bounces, oscillating bounces were discussed in the literature \cite{Hackworth:2004xb,Lee:2011ms}. These provide a bridge between CdL and HM instantons. For solutions with $n$ oscillations these were shown to contain $n$ negative modes in the spectrum of linear perturbations in the homogeneous sector \cite{Lavrelashvili:2006cv,Battarra:2012vu}, which strongly suggests that only solutions with a single interpolation have physical relevance.

\begin{figure}[ht!]
	\centering
	\includegraphics[width=0.3\textwidth]{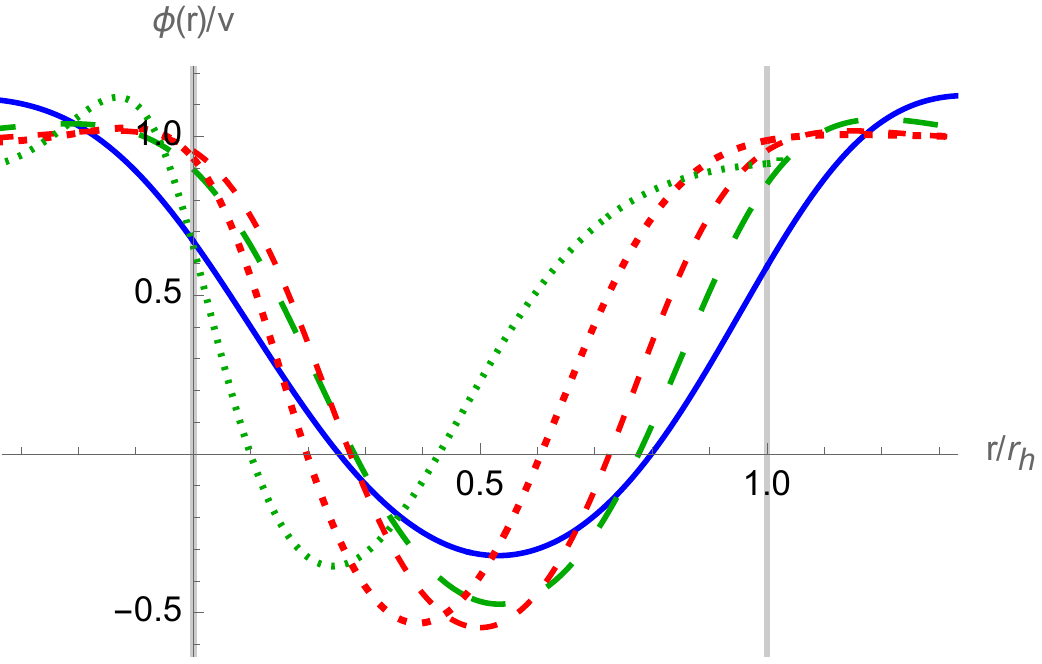} \,\, \includegraphics[width=0.3\textwidth]{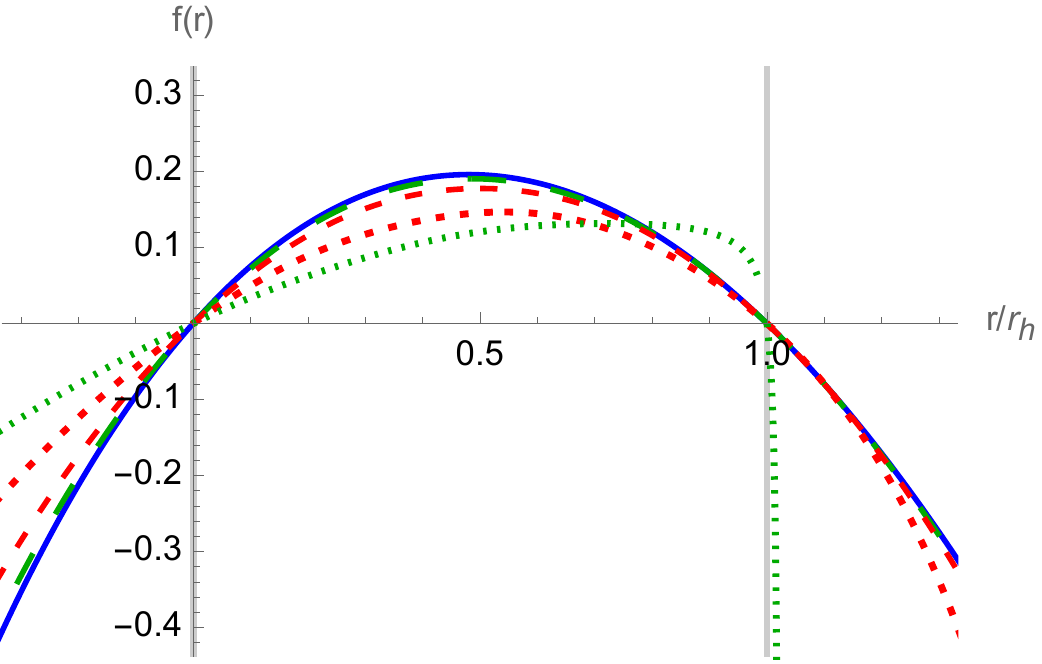} \,\, \includegraphics[width=0.3\textwidth]{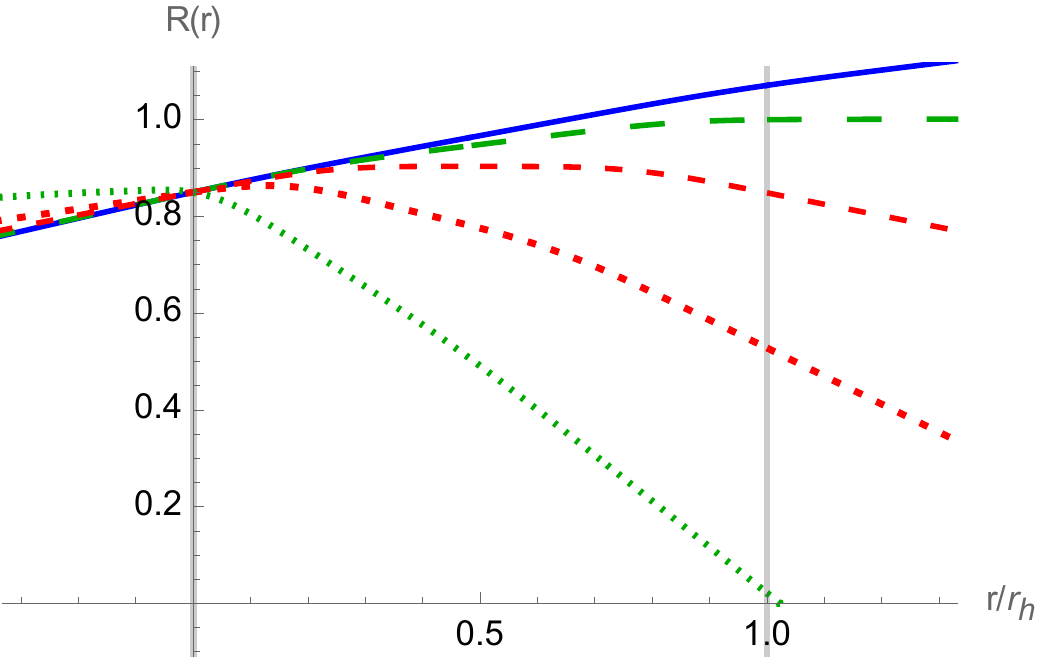}
	\caption{Series with $n=2,$ i.e. two interpolations of the scalar field across the potential barrier. At $\lambda=5000$ and for one horizon of size $R_1=0.85,$ with values $v=0.04,0.049,0.0618,0.08,0.1$ corresponding to curves that are progressively more finely dashed. Note that the $r$ coordinate has in each case been rescaled such that the horizons lie at $r=0,1$ while the scalar field value is shown w.r.t. the values $\pm v$ at the potential minima.}
	\label{fig:n2}
\end{figure}

The influence of primordial black holes on vacuum decay was questioned long ago \cite{Hiscock:1987hn,Berezin:1987ea,Berezin:1990qs} and received renewed interest recently \cite{Gregory:2013hja,Burda:2015isa,Burda:2015yfa,Burda:2016mou,Gregory:2020cvy}. These authors considered static oscillating bounces interpolating between a seed black hole horizon (BHH) and a de Sitter cosmological horizon (CH), i.e. precisely the kind of solution which we have studied (and generalised) in the present paper. Such solutions are conjectured to act much like impurities in condensed matter, with black holes acting as catalysts for phase transitions by being the preferred nucleation sites for bubbles of true vacuum.

\begin{table}[ht!]
\begin{tabular}{ |p{3cm}|p{3cm}|p{3cm}|p{3cm}|p{3cm}|}
		\hline
		$v$ & \textbf{$\varphi_1$}& \textbf{$\varphi_2$} & \textbf{$f'_2$} & \textbf{$R_{2}$} \\
		\hline
		0.040 & 0.0268 & 0.0236 & 0.858 & 1.07 \\
		\hline
		0.049 & 0.0440 & 0.0417 & 0.902 & 1.00 \\
		\hline
		0.0618 & 0.0590 & 0.0590 & 1.00 & 0.85 \\
		\hline
		0.080 & 0.0746 & 0.0790 & 1.26 & 0.528 \\
		\hline
		0.100 & 0.0656 & 0.0914 & 18.1 & 0.021 \\
		\hline
		\end{tabular}
\caption{Optimised horizon values for the solutions in Fig. \ref{fig:n2}, with $\lambda=5000$ and fixed $R_{1}=0.85.$}\label{tablen2}
\end{table}

Let us mention first that in our setting one may also find solutions in which the scalar field oscillates several times across the potential barrier. Such solutions typically arise only at larger values of the self-coupling $\lambda,$ as suggested for instance by the limits of existence of symmetric solutions in Eq. \eqref{eq:Nariailimit}. Examples of solutions with $n=2$ interpolations are given in Fig. \ref{fig:n2}, and their corresponding optimised parameters in table \ref{tablen2}. Again we find solutions of all types: blue solutions in which the size of the transverse sphere changes monotonically, and which asymptote to de Sitter spacetime outside of the cosmological horizon. Then there are green solutions, in which the cosmological horizon hides a collapsing universe. Finally, there are red solutions which correspond to oscillating scalar lumps surrounded by two black holes. It would be straightforward to search for solutions with more interpolations of the scalar field, but by analogy with oscillating bounces we may assume that, for $n$ oscillations, such solutions contain $n$ negative modes in their spectrum of linearised homogeneous perturbations \cite{Battarra:2012vu}, and are thus not dominant in the process of vacuum decay. Hence, from here on, we will restrict our discussion to solutions with a single interpolation.

\begin{figure}[ht!]
	\centering
	\includegraphics[width=0.5\textwidth]{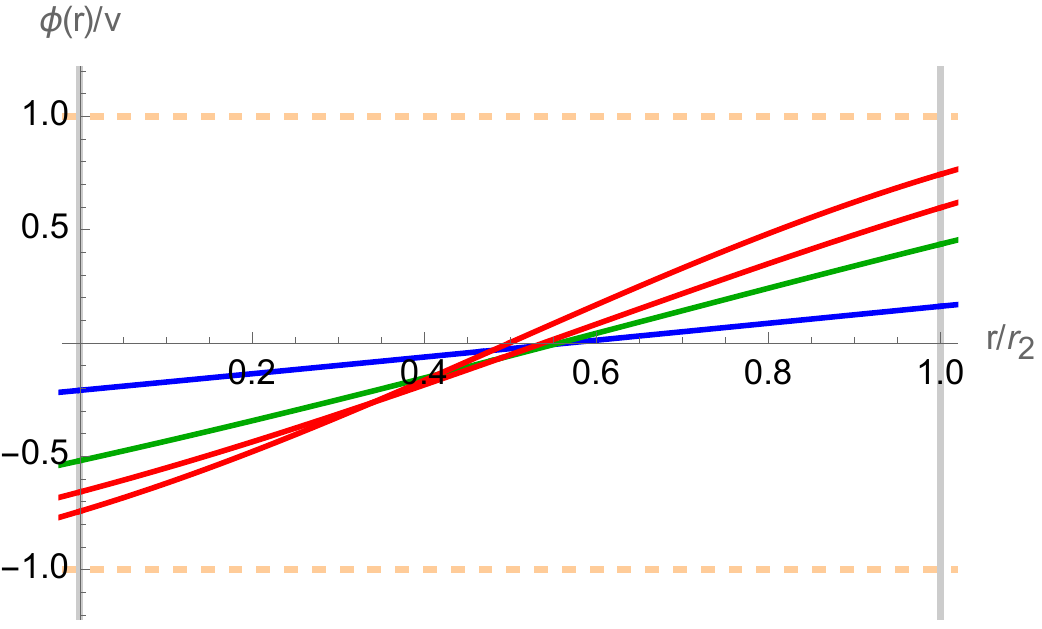}
		\caption{The scalar field evolution for solutions with $v=0.18, R_1=0.6$ for the series $\lambda=115,150,200,295.$ This figure illustrates that the potential minima are not reached at the horizons.}
	\label{fig:interpol}
\end{figure}

It is useful to first examine the behaviour of the scalar field in more detail. For this reason we have provided a magnified version of the solutions from Fig. \ref{fig:examples}, where we only depict the evolution in between horizons -- see Fig. \ref{fig:interpol}. As is immediately evident, the scalar field does not interpolate between the minima of the potential, but rather stays up on the barrier for the entire evolution. This is in direct analogy with CdL bounces, which also do not start/end at the potential minima. This peculiar feature of gravitational instantons is worth examining in more detail.

To contrast with the current case, first recall that in flat space the bounce equation is
\begin{align}
\frac{d^2\varphi}{d\xi^2}+\frac{3}{\xi} \frac{d\varphi}{d\xi}-\frac{dV}{d\varphi}=0 \kma
\end{align}
where $\xi = \sqrt{\tau^2+r^2}$. The bounce solution starts at some point $\varphi=\varphi_0$ at $\xi =0$ and we would like to see if it can approach a local minimum (false vacuum), where $dV/d\varphi=0,$ as $\xi \to \infty.$ Let's denote by $\delta \varphi$ a deviation of the scalar field from the false vacuum value,
\begin{align}
\varphi (\xi) = \varphi_f + \delta \varphi (\xi) \kma
\end{align}
then neglecting the friction term for very late times, we get
\begin{align}
\left(\frac{d^2}{d\xi^2}-\mu^2\right)\delta\varphi = 0 \kma
\end{align}
where $\mu^2 = \frac{d^2 V}{d^2 \varphi}|_{\varphi=\varphi_f}$
And we see that the solution approaches the false vacuum exponentially fast as $\xi \to \infty$,
\begin{align}
\varphi (\xi) = \varphi_f + e^{-\mu\xi} \pkt
\end{align}

The situation changes when gravity is included. For CdL solutions, the relevant scalar field equation is
\begin{align}
\frac{d^2\varphi}{d\xi^2}+\frac{3 \dot{\rho}}{\rho} \frac{d\varphi}{d\xi}-\frac{dV}{d\varphi}=0 \kma
\end{align}
where $\rho(\xi)$ is the Euclidean scale factor and $\xi$ now varies in a finite interval $\xi = [0, \xi_{max}]$
while $\rho$, in addition to the origin, develops a second zero at some $\xi = \xi_{max}$.
Assuming that the bounce ends at the false vacuum, $\varphi \to \varphi_f$ as   $\xi \to \xi_{max}$,
close to the false vacuum we get the equation
\begin{align}
(\frac{d^2}{d\rho^2}+\frac{3}{\rho} \frac{d}{d\rho}-\mu^2)\delta\varphi = 0 \kma
\end{align}
where $\rho= \xi_{max}-\xi$. The general solution of this equation is given in terms of two Bessel functions
\begin{align}
\delta\varphi (\mu \rho) = C_1 \frac{K_1(\mu\rho)}{\mu \rho} + C_2 \frac{I_1 (\mu\rho)}{\mu\rho} \pkt
\end{align}
We see that the assumption of a small deviation is not fulfilled,
since one function is divergent and the other goes to constant as $\rho \to 0$,
\begin{align}
\frac{K_1(\mu\rho)}{\mu \rho} \to \frac{1}{\rho^2}, ~~~ \frac{I_1 (\mu\rho)}{\mu\rho} \to \frac{1}{2} \pkt
\end{align}
We conclude that the bounce cannot end at the false vacuum.

For our scalar lump solutions we find an analogous situation, if we consider the endpoints of the solutions to reside at the horizons.
Let us assume then that the scalar lump solutions reach the false vacuum at a horizon. To linear order we then obtain (see also Eq. \eqref{scalarpert})
\begin{align}
\left(\frac{d^2}{dy^2}+\frac{1}{y} \frac{d}{dy}- \frac{a^2}{y}\right)\delta\varphi (y) = 0 \kma
\end{align}
where $y=r_c-r$ and $a^2 \propto \mu^2 >0$.
The general solution to this equation is also given in terms of two Bessel functions,
\begin{align}
\delta\varphi(y) = C_1 K_0(2 a \sqrt{y}) + C_2 I_0 (2 a \sqrt{y}) \pkt
\end{align}
We see that the assumption of a small deviation from the false vacuum is again unjustified,
since one Bessel function is divergent and the other goes to a constant as $y \to 0$,
\begin{align}
K_0(2 a \sqrt{y}) \to -\ln(a\sqrt{y}), ~~~  I_0 (2 a \sqrt{y}) \to 1 \pkt
\end{align}
We again arrive at a contradiction and conclude that static oscillating bounces cannot end
exactly at the false vacuum at the location of the horizons, contrary to what is claimed in~\cite{Gregory:2020cvy}. In addition, since the scalar field cannot reach a potential minimum at the horizon, its first derivative must also be non-zero there, as implied by the expansion \eqref{eq:phiTaylorBHgg}. Note that beyond the horizon, the scalar field may well undergo damped oscillations around the false vacuum and thus reach the false vacuum at future infinity. However, for the purposes of vacuum decay, the idea (used in \cite{Gregory:2020cvy}) is to glue a scalar lump to an outside de Sitter (or SdS) solution at the location of the cosmological horizon. We now see that this cannot actually be done -- the scalar field cannot continuously be matched.

We can see two ways in which the obstruction just identified could be resolved. Before discussing them, let us mention though that the situation is in fact worse than previously described: it is not just the scalar field that behaves non-smoothly. At the horizon, by definition the metric function $f$ is zero and can thus be matched continuously. Also, since in a fixed potential scalar lump solutions arise as a one-parameter family of solutions with varying horizon sizes, one may choose the horizon size to match onto the desired ``outside'' de Sitter (or SdS) spacetime. However, neither $f^\prime$ nor $R^\prime$ can in general be matched continuously. This means that in general the gluing of a scalar lump instanton to an outside SdS spacetime has discontinous values of $\varphi, \varphi^\prime, f^\prime$ and $R^\prime.$

One interpretation might be that the gluing is not complete yet. A second quantum transition may be needed to match the discontinuous field values to each other. This might then considerably lower the transition rate, and may make black hole seeded vacuum decay less likely than claimed. If no such solution may be found, then the transition would in actuality correspond to an off-shell transition, which would be vastly more suppressed still. We note that a discontinuity in $R^\prime$ at the horizon might not be penalised by a larger action, since all terms involving $R^\prime$ in the action \eqref{eq:redact} are proportional to $f,$ which vanishes at the horizon. However, discontinuities in $f^\prime$ do not share this property, and must be taken into account.

Another possibility is to interpret the jump in field values as due to thermal activation of the instanton transition. This is the interpretation proposed for CdL bounces in \cite{Brown:2007sd}. There, the authors considered the surrounding de Sitter spacetime as a heat bath, from which the scalar field can jump up part of the potential barrier due to a thermal fluctuation, and then complete the transition across the barrier via the CdL bounce. It is conceivable that a similar interpretation is possible here. However, to make this analysis precise and truly convincing, one would have to go beyond the treatment in \cite{Brown:2007sd} and include the backreaction of the spacetime. As we saw above, there are necessarily discontinuities in the background geometry and these would have to be addressed within a more complete framework. This interesting question goes beyond the aims of the present paper. For now, we may conclude that more work is required to fully understand black hole seeded vacuum decay. \footnote{For a related, but different, approach see also \cite{Shkerin:2021zbf}.}


\section{Concluding remarks}

We have investigated solutions of general relativity exhibiting two horizons, in the presence of a scalar field with a potential barrier. In addition to known solutions, for which the size of transverse 2-spheres grows indefinitely,
we have found new classes of solutions, for which the size of transverse 2-spheres shrinks at
some distance away from the first horizon, either after the second horizon or even before it. These new solutions exist over large regions of the parameter space of the potential.

\begin{figure}[ht!]
	\centering
	\includegraphics[width=0.3\textwidth]{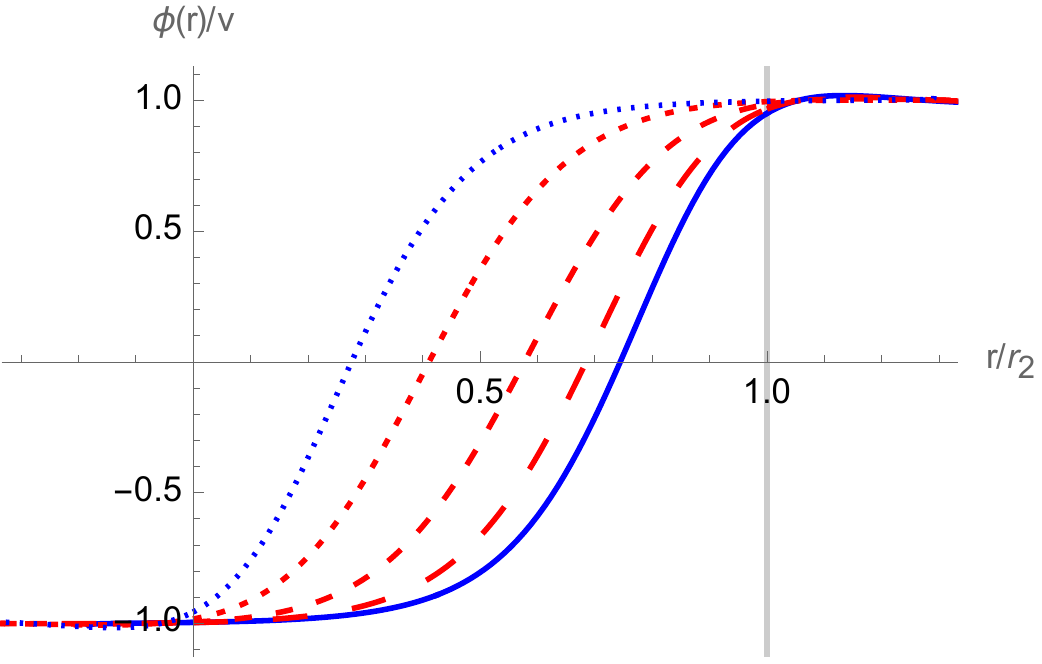} \,\, \includegraphics[width=0.3\textwidth]{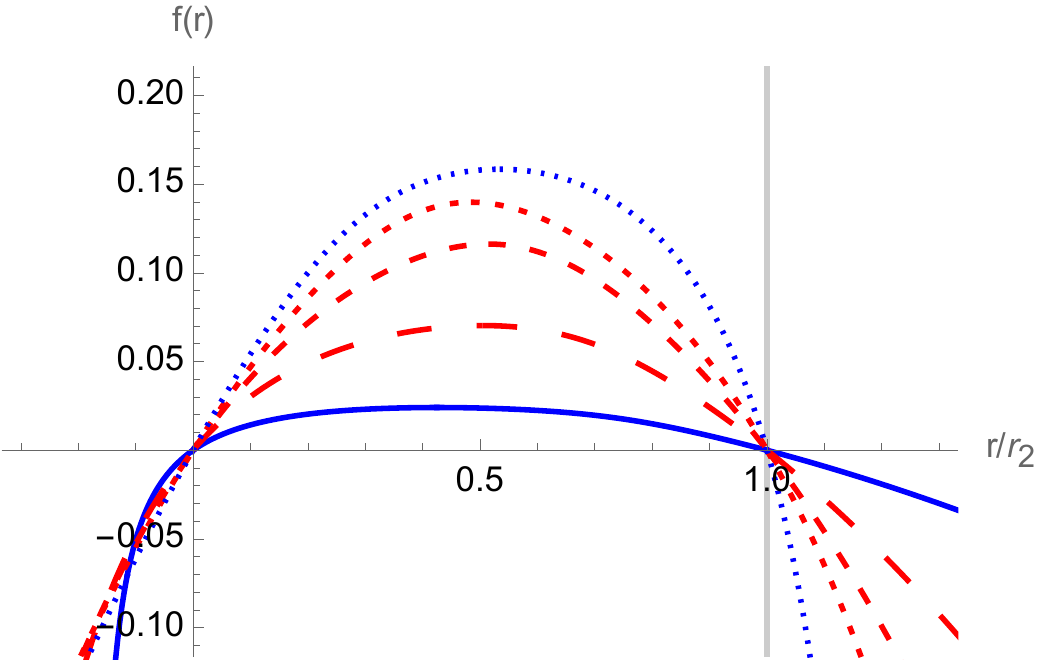} \,\, \includegraphics[width=0.3\textwidth]{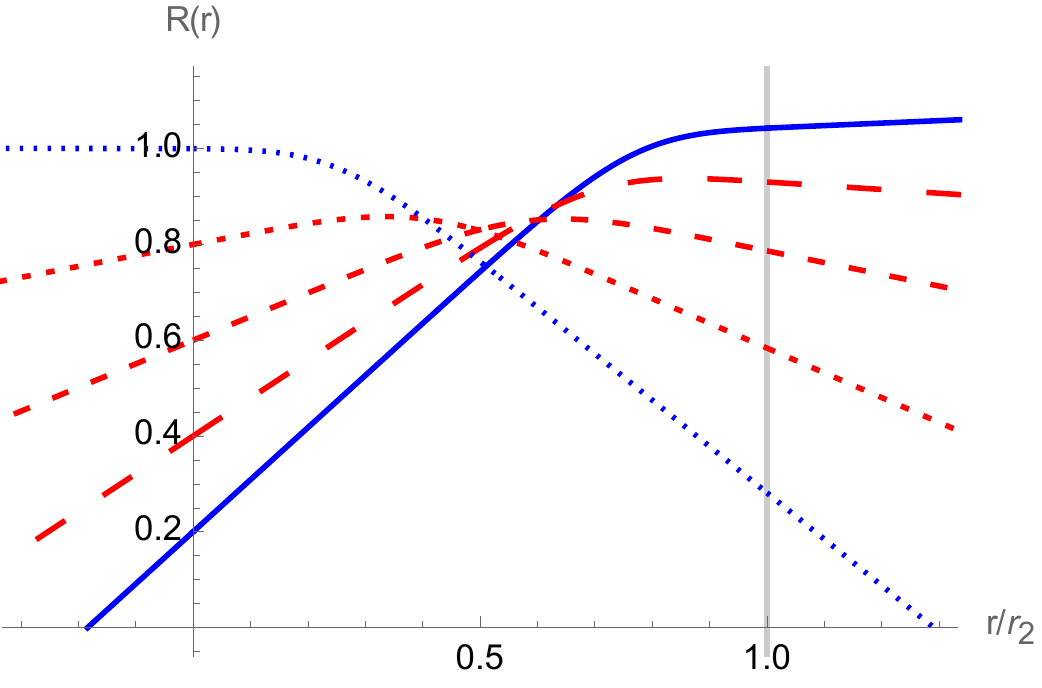}
	\caption{Examples of various solutions at fixed potential, $\lambda=2000, v=0.11.$ Increasingly fine dashing corresponds to a larger first horizon size, with solutions shown for $R_1=0.2,0.4,0.6,0.8,1$. For a description, see the main text.}
	\label{fig:horizonseries}
\end{figure}

A few words on the physical interpretation of these solutions, see also Fig. \ref{fig:horizonseries}: if we start with the solutions containing a single horizon, then we may describe them as a spherical wall of scalar energy that is prevented from collapsing under its own weight due to the vacuum energy it encloses. Put differently, the wall has a tendency to collapse due to its tension, but the spacetime inside it has a compensating tendency to expand due to the positive vacuum energy it contains, and an equilibrium can be reached. At equilibrium, the wall resides at a radial location that is close to the horizon. Now we add a black hole at the centre of this configuration. This black hole increases the gravitational pull on the scalar domain wall. As a result, the wall moves closer to the black hole horizon, and this effect is enhanced as the black hole grows. At fixed parameter values for the potential, as the first horizon grows, the second horizon shrinks until a symmetric solution is reached. For this symmetric solution the wall has moved precisely to the middle of the two horizons, which are now both representing black holes. As the first horizon is further increased, we recover reflected versions of the solutions described before (up to an irrelevant overall scaling of $f(r)$), with the two horizons being of unequal sizes. In all cases, the wall sits closer to the larger of the two horizons. 

One question one may ask is how realistic such solutions might be. Here it is interesting to note that in our current understanding of fundamental physics, potentials with a barrier are highly plausible: in particular, the current understanding is that the Higgs potential is effectively of this type, given current measurements of the top quark mass \cite{Degrassi:2012ry}. Moreover, scalar potentials with a barrier are also conjectured to be common in string theory, since they can evade the de Sitter swampland conjectures \cite{Garg:2018reu,Ooguri:2018wrx}. Conceivably, scalar lumps may thus impact the history of the universe, both at early and late times. Since scalar lump solutions interpolate between different sides of a potential barrier, at early times they may have played a role in determining the coupling constants of our universe, given that in string theory all coupling constants arise as expectation values of scalar fields. In the late universe, a question of interest is that of the stability of our current vacuum. Scalar lumps supported by the Higgs, or other scalar fields, may cause our vacuum to destabilise near black holes, if indeed vacuum decay can be catalysed by black holes. It is thus of evident interest to study black hole seeded decay in more detail, especially given the open questions discussed in the previous section.

As a final pointer for future work, let us remark that similar static oscillating solutions were also studied for Yang-Mills fields in \cite{Volkov:1996qj}. In the same vein, it would then be interesting to see what bearing such solutions might have on the problem of vacuum stability.

\acknowledgments

We are thankful to Sebastian Bramberger for collaboration at the initial stage of this project.
The work of G.L. is supported in part by the Shota Rustaveli National Science Foundation of Georgia with Grant N FR-19-8306. JLL gratefully acknowledges the support of the European Research Council (ERC) in the form of the ERC Consolidator Grant CoG 772295 ``Qosmology.''


\end{document}